\documentstyle[epsfig,amsfonts,amssymb,diagram,preprint,tighten,floats,aps]{revtex}

\newcommand{\R}{{\Bbb R}}
\newcommand{\C}{{\Bbb C}}
\newcommand{\Z}{{\Bbb Z}}
\newcommand{\Sch}{{\mathfrak S}} 

\def\conf{{\rm SL}(2,\C)}
\def\isom{{\rm SL}(2,\R)}

\def\H{{\bf H}}
\def\S{{\bf S}}

\def\Q{{\cal Q}}
\def\a{\alpha}
\def\b{\beta}
\def\N{{\cal N}}
\def\w{\omega}

\begin{document}

\title{Holography and Riemann Surfaces}

\author {Kirill Krasnov}

\address{Physics Department, University of California,\\
Santa Barbara, CA 93106, USA} 

\maketitle

\begin{abstract}

We study holography for asymptotically AdS spaces with an arbitrary 
genus compact Riemann surface as the conformal boundary. Such spaces can be 
constructed from the Euclidean ${\rm AdS}_3$ by discrete 
identifications; the discrete groups one uses are the so-called 
classical Schottky groups. As we show, the spaces so constructed 
have an appealing interpretation of ``analytic continuations'' of 
the known Lorentzian signature black hole solutions; it is one 
of the motivations for our generalization of the holography to this case.
We use the semi-classical approximation to the gravity path integral, and
calculate the gravitational action for each space, which is given by  
the (appropriately regularized) volume of the space. As we show, the
regularized volume reproduces exactly the action of Liouville theory, 
as defined on arbitrary Riemann surfaces by Takhtajan and Zograf. 
Using the results as to the properties of this action, we discuss thermodynamics of the 
spaces and analyze the boundary CFT partition function. Some aspects 
of our construction, such as the thermodynamical interpretation of 
the Teichmuller (Schottky) spaces, may be of interest for mathematicians 
working on Teichmuller theory.

\end{abstract}

\eject

\section{Introduction}
\label{sec:intr}

In this paper we apply the idea of bulk/boundary correspondence to a
particular large class of asymptotically AdS spaces. More precisely, 
we study holography in 2+1 dimensions, and the spaces we
consider have non-trivial boundary topology. As we show, the holographic 
prescription can be meaningfully extended to the case of manifolds with 
any Riemann surface as the conformal boundary.

We wish to emphasize from the outset that the paper is ``classical''
in its spirit - no stringy aspects of the idea of holography
is used. Our analysis of the boundary 
CFT is based on the semi-classical approximation to the gravity path 
integral. Thus, our results will pertain, in the regime of validity of 
this approximation, to any quantum gravity theory in 2+1 dimensions, 
in case any such consistent theory other than string theory is ever 
constructed in the future. 

The idea to study holographically dual CFT's for arbitrary topology
of the boundary can be suggested by purely conformal field theory 
considerations. Indeed, two dimensional CFT's certainly do make sense 
on any genus Riemann surface. In fact, any CFT can be viewed as exactly the 
structure that endows the moduli space of Riemann surfaces with 
analytic geometry \cite{FS}. Moreover, knowing the CFT partition function for
arbitrary genus allows one to reconstruct all correlation functions.
This is done by varying the moduli so as to 
``pinch'' handles of the Riemann surface; the correlation functions
can then be read off from the leading order terms of the expansion
in the moduli, see \cite{FS}. Thus, from the CFT point of view, it
is quite natural to allow the genus and moduli to be variable.

Another CFT reason to consider different topologies is the fact that
certain vertex operators can ``change'' the topology of the surface
where the theory lives. These are the multi-valued operators, such
as fermionic vertex operators or certain operators in orbifold 
constructions. When one calculates correlation functions of such
operators, it is very convenient, and sometimes necessary, to
introduce the universal cover of the worldsheet. One then gets
Riemann surfaces of various genera; the classical fact that a Riemann surface
of genus $g$ covers the sphere $g+1$ times is a good illustration.
See \cite{Vafa} for various examples.

Last, but not least of our motivations, is that, as is demonstrated
in the next section, asymptotically AdS spaces with non-trivial boundary 
topology are, in certain precise sense, the ``analytic continuations'' to 
the Euclidean signature of the well-known black hole solutions. A 
particularly interesting set of black hole solutions consists of
black holes with a single asymptotic region and an arbitrary genus
wormhole contained inside the horizon. As we show in the next section,
the boundary of the Euclidean counterparts of these spacetimes is
a Riemann surface, whose genus is twice the genus of the wormhole.
Thus, extending the ideas of holography to Euclidean spaces of
non-trivial boundary topology is a step towards the holographic
description of these black hole spacetimes. Potentially, this gives
a way towards understanding whether the holography can 
``see'' the part of spacetime hidden behind the horizon.

Let us now, for the sake of readers who are not familiar with it, 
review the very basics of the idea of holography, in the
amount that is needed for the purposes of this paper. 
In broad terms holography is a description of a theory in $n+1$
dimensions by another theory in $n$ dimensions. Numerous examples
of such a description were recently found in string theory, see
\cite{String:review} for a review. 
The holographic prescription of \cite{GKP,Witten-Hol} states that
the generating functional of the boundary field theory is
given by the bulk path integral, with sources being just the 
(appropriately rescaled) boundary values of the bulk fields. 
If one is only interested in correlation functions of the
boundary stress-energy tensor, one should put sources for
all other boundary operators to zero. Thus, the generating 
functional of correlation functions of the stress-energy 
tensor (or CFT partition function) is given by the bulk gravity path integral.
The source is just the (appropriately rescaled) boundary
value of the metric, and the path integral sums
over all bulk metrics approaching this fixed boundary value.
The semi-classical approximation to this path integral
is given by the sum over all classical solutions of
Einstein equations that approach a fixed metric at the
boundary.

Thus, to find the semi-classical approximation to the boundary
CFT partition function, we have to identify all classical
solutions that approach a prescribed metric at
the boundary. This is done in Section \ref{sec:manifolds}, where
we describe Euclidean spaces of constant negative curvature whose conformal
boundary is an arbitrary Riemann surface. These are classical results, 
dating back to the work of Poincare on Kleinian groups. 
In this section we also discuss the ``spacetime'' interpretation of
the Euclidean spaces constructed. We show that in certain precise sense they
are ``analytic continuations'' to Euclidean signature of physically 
interesting black hole solutions. The CFT partition function is then 
studied in Section \ref{sec:part}. Here we describe the regularization 
procedure that is needed to render the on-shell gravity action finite, discuss the
thermodynamics of our Euclidean spaces, and analyze the resulting CFT
partition function. We conclude with a discussion.

An important role in this paper is played by the results of 
Takhtajan and Zograf \cite{Takht} on the Liouville theory on 
Riemann surfaces. We ``derive'' the action proposed in \cite{Takht} using the
bulk/boundary prescription. Thus, this paper gives a new, (2+1)-dimensional
interpretation to the results that were obtained based on purely 
two-dimensional considerations. Moreover, our observation that the
Euclidean spaces considered can be interpreted as the 
``analytic continuations'' of the black hole spacetimes
allowed us to give the results of \cite{Takht} a 
thermodynamical interpretation. 

Most of the mathematics background of this paper, including the
definition of the Liouville action for any Riemann surface, and
results on the dependence of this action on the moduli, is from 
work \cite{Takht}. To make the paper self-consistent, some of this 
material is repeated here. Thus, the relevant facts on Schottky groups
are reviewed in \ref{ss:euclid}, and some facts about Liouville theory
on Riemann surfaces are reviewed in \ref{ss:unif}. The Appendix
reviews aspects of the theory of Teichmuller spaces, 
which play an important role in our story.

There exists a rather large literature on bulk/boundary correspondence
in 2+1 dimensions and we would like to mention few papers
that are somewhat related to the present work. Thus, the correspondence for the
boundary topology different from that of a sphere was discussed
in \cite{Bonelli,WY}. The genus one case was discussed in
\cite{Birmingham,Chekhov}. A relation between geometry of hyperbolic
manifolds and the boundary CFT was discussed in \cite{Kholodenko}.
Finally, the observation that the Liouville action appears as
the effective action on the boundary was made in 
\cite{Skenderis,D1D5,Bautier}.

\section{Classical solutions}
\label{sec:manifolds}

This section consists of two parts. Subsection \ref{ss:euclid} describes
Euclidean Einstein manifolds of constant negative curvature that 
have a Riemann surface as the boundary. The spaces of interest
are well-known since the work of Poincare on Kleinian groups. These are
the spaces obtained by discrete identifications of AdS${}_3$ using 
Schottky groups. All facts discussed in this subsection are well-known.
Subsection \ref{ss:lor} discusses the Lorentzian signature interpretation 
of the spaces. This interpretation is, to the best of our
knowledge, new.

\subsection{Euclidean case}
\label{ss:euclid}

In three dimensions all constant negative curvature Einstein manifolds are
locally indistinguishable from the AdS space. Thus, they can be obtained 
by an appropriate identification of points in AdS. 

In what follows, we shall mostly use
the upper half-space model of AdS. We denote the upper half-space by
$\H$, which stands for the `hyperbolic space'. 
The hyperbolic metric on $\H$ is:
\begin{equation}
ds^2 = {l^2\over z^2}(dx^2+dy^2+dz^2),
\end{equation}
where $l$ is the curvature radius. The group of isometries of $\H$ is $\conf$. 
The boundary $\S$ of $\H$ is at $z=0$, plus the point at infinity. 
Thus the boundary is just the Riemann sphere, and 
$\conf$ acts on $\S$ by conformal transformations.
This action of $\conf$ on $\S$ can be described most elegantly by
introducing the complex coordinate $w=x+iy$. Then the action of $\conf$ on 
$\S$ is given by fractional linear transformations 
(or {\it Mobius} transformations):
\begin{equation}
w \to {aw+b\over cw+d}, \qquad a,b,c,d \in \C, \qquad ad-bc =1.
\end{equation}
The action of $\conf$ on $\H$ itself can also be represented in terms of
fractional linear transformations; this is done by introducing quaternionic
coordinates in the upper half-space, see, e.g., \cite{Hyperb-Geom} for 
details of this construction.

In what follows we will need some basic facts about the $\conf$ action in $\H$. 
The geodesics are semi-circles that intersect the boundary $\S$
orthogonally (lines orthogonal to $\S$ included). 
Transformations from $\conf$ map geodesics to geodesics. The geodesic
surfaces of $\H$ are hemispheres intersecting the boundary orthogonally
(the planes orthogonal to $\S$ included). 
Geodesic surfaces are mapped to geodesic surfaces. 
An important fact about the action of $\conf$ is that each such 
transformation can be represented by an even number of inversions
in hemispheres in $\H$. 

One can obtain constant negative curvature spaces other than $\H$ 
identifying its points by the action of some discrete subgroup 
of $\conf$. To describe the action of such a group 
it is very convenient to concentrate on its action on the boundary. 
Let $\Delta$ be the 
(closure of the) set of fixed points of this action. Thus, by
definition $\Delta$ is a closed set. If it does not coincide with
the whole of $\S$, the group is called a {\it Kleinian} group. 
We denote Kleinian groups by $\Sigma$.
Then $\Omega=\S\backslash\Delta$ is an open set 
on which $\Sigma$ acts discontinuously. It is called
the {\it region of discontinuity} of $\Sigma$. The quotient $X=\Omega/\Sigma$ is
a smooth manifold. For what follows we will also need the notion of
a {\it fundamental region} for $\Sigma$. It is a 
set $D\subset\Omega$, such that no two distinct interior points 
of $D$ are $\Sigma$-equivalent, and every point of $\Omega$ is 
$\Sigma$-equivalent to some point of $D$. 
We shall see examples of fundamental
regions below. For more information on Kleinian groups see,
e.g., \cite{Kleinian}. Extending the action of $\Sigma$ to the whole $\H$, and
identifying points that are $\Sigma$-related, one
gets a space of constant negative curvature whose boundary is $X$.

Varying $\Sigma$ one gets a plethora of spaces, some of them of complicated
topology. In this paper we shall consider only rather special Kleinian
groups, known as {\it classical Schottky groups}. These are the 
simplest, and most studied examples of Kleinian groups. They were
first discovered by Schottky and Klein, and then studied in detail
by Poincare more than a hundred years ago. As we shall see in a moment, 
it is these groups that lead to compact Riemann surfaces
as boundaries. Thus, they are general enough for the purpose of this
paper, which is to study the holography on manifolds bounded by
Riemann surfaces. Unlike the spaces one gets for more
complicated Kleinian groups, the spaces associated with the
classical Schottky groups
also admit a direct Lorentzian interpretation, as we shall see in the
next subsection. We only consider classical Schottky groups in this
paper.

A Schottky group $\Sigma_g$ of genus $g$ is (freely) 
generated by a finite number $g$ of
loxodromic generators $L_1,\ldots,L_g$ (a group element
is called loxodromic if it is conjugate, in the group of all
Mobius transformations, to a transformation of the form
$w\to w/\lambda, |\lambda|>1$). 
The number $g$ of generators of a Schottky group $\Sigma$ is exactly
the genus of the resulting quotient space $X$.
We also need the
notion of a {\it marked} Schottky group. A Schottky group 
is called marked if the set of its generators is ordered.

A marked Schottky group is most conveniently 
described by its fundamental region. Let 
$C_1,\ldots,C_g,C_1',\ldots,C_g'$ be $2g$ non-intersecting
circles in $\S$ such that all circles
lie to the exterior of each other. Let $L_i$ be the loxodromic
element mapping $C_i$ to $C_i'$ so that the region
exterior to $C_i$ is mapped to the interior of $C_i'$. Then the group $\Sigma_g$
freely generated by the elements $L_1,\ldots,L_g$ is a Schottky group,
its fundamental region is the part of $\S$ exterior to all circles,
and the quotient space is a compact Riemann surface of genus $g$.
To convince oneself that the quotient space obtained this way
is indeed a genus $g$ Riemann surfaces it is best to picture the boundary as
$S^2$. Cutting out of this sphere $2g$ discs one gets a manifold with
a boundary, the boundary being just the collection of circles 
$C_1,\ldots,C_g,C_1',\ldots,C_g'$. One then identifies the circles
to obtain a manifold without boundary, which is just a $g$-handled sphere.

Using the freedom of conjugating all generators by some Mobius transformation,
one can choose the fundamental region in the canonical way. Each generator
$L_i$ is completely characterized by its two fixed points (because all
$L_i$ are strictly loxodromic these points are distinct), and the value
of its {\it multiplier} $\lambda_i$. The geometrical meaning of the 
multiplier is to specify by ``how much'' the corresponding
transformation moves things along the geodesics connecting the fixed points.
Let us denote the attractive fixed point of $L_i$ by $a_i$ and its
repulsive fixed point by $b_i$. Then, by conjugation in the group
of all Mobius transformations, we can put $a_1=0, b_1=\infty, a_2=1$.
A Schottky group for which these conditions hold is called normalized.

Let $\Sigma_g$ be a marked Schottky group of genus $g$. The map
\begin{equation}
\Sigma_g \to 
(a_3,\ldots,a_g,b_2,\ldots,b_g,\lambda_1,\ldots,\lambda_g) \in \C^{3g-3}
\label{coord-schottky}
\end{equation}
establishes a one-to-one correspondence between the set of marked 
Schottky groups and some subset of $\C^{3g-3}$. This subset can
be shown to be a connected region, and is called the {\it Schottky space}
of genus $g$. We will denote it by $\Sch_g$.
Because of its dimension (real 6g-6) one is tempted to think
that the Schottky space $\Sch_g$ is related in some way to the moduli space
of Riemann surfaces of genus $g$. This is indeed the case, the moduli
space $R_g$ turns out to be a certain quotient of $\Sch_g$, as we
shall explain below.

The classical {\it retrosection theorem} due to Koebe states that
every compact Riemann surface $X$ can be represented as the quotient
$\Omega/\Sigma$, where $\Sigma$ is a Schottky group with region of
discontinuity $\Omega$. Moreover, one can always choose a
Schottky group and the fundamental region in such a way that a given 
set of $g$ generators of $\pi_1(X)$ is the image of circles $C_i$
under the quotient map. More precisely, given a basis of generators
\begin{equation}
\a_1,\ldots,\a_g,\b_1,\ldots,\b_g: \prod_{i=1}^g [\a_i,\b_i] = 1, \qquad
[\a,\b]=\a^{-1}\b^{-1}\a\b
\label{Fuchs-rel}
\end{equation}
of $\pi_1(X)$, one can choose a marked Schottky group $\Sigma_g$ and $g$ 
generators $L_1,\ldots,L_g$ of $\Sigma_g$ so that there is a standard 
fundamental region $D$ bounded by $C_1,\ldots,C_g,C_1',\ldots,C_g'$ 
with $L_i(C_i)=C_i'$ such that: (i) $\a_i$ is the image of $C_i$ under the 
quotient map $\Omega\to\Omega/\Sigma_g$; (ii) the group of the
covering $\Omega\to\Omega/\Sigma_g$ coincides with the smallest
normal subgroup $\N$ of $\pi_1(X)$ containing elements $\a_1,\ldots,\a_g$;
(iii) group $\Sigma_g$ is isomorphic to the factor group $\pi_1(X)/\N$.
A proof of this theorem can be found, e.g., in \cite{Aut}. For more
information on Schottky groups see \cite{Takht} and 
references therein.

Thus, all compact Riemann surfaces arise as quotients of $\S$ by
classical Schottky groups. This fact allows one to use the space
of all Schottky groups --the Schottky space-- to coordinatize
the space of Riemann surfaces. The problem is that several
points in the Schottky space may correspond to the same 
Riemann surface. This turns out to be indeed the case, see
\cite{Sch-R}. Thus, to get from $\Sch_g$ to the space of Riemann 
surfaces $R_g$ one performs certain identifications. This is discussed 
in more details in the Appendix.

The procedure of associating a compact Riemann surface with a Schottky
group such that the Riemann surface is represented as the 
quotient space is called {\it uniformization} by Schottky groups.
There is another well-known uniformization procedure --the so-called
uniformization by Fuchsian groups-- that we will encounter below.

To summarize, we see that one can get constant negative curvature
manifolds whose boundary is a Riemann surface by doing identifications
of points in AdS${}_3$ by the action of a Schottky group. Moreover,
the retrosection theorem guarantees that {\it all} Riemann surfaces arise
as boundaries of the spaces so constructed.

Let us now illustrate this construction on the examples of $g=1,2$.

\bigskip
\noindent{\bf Example A: g=1}

The genus one case is different from all higher genus cases, for it is
uniformized by the zero curvature space, while all higher genus
surfaces are uniformized by the space of negative curvature. Also, the
(real) dimension of its moduli space (which is two) does not follow 
the $6g-6$ formula of the higher genus cases. Still, it 
gives a good illustration of the general construction. It is also
of direct physical relevance, for the corresponding space describes,
e.g., the AdS space at finite temperature and BTZ black hole.
 
The genus one Schottky groups are generated by a single loxodromic
element $L$. Thus, elements of such a group are simply powers of the
generator (and its inverse). The set $\Delta$ of fixed points
consists in this case of only two points: the attractive and
repulsive fixed points of the element $L$. This is an example
of the so-called elementary Schottky groups.

\begin{figure}
\centerline{\hbox{\epsfig{figure=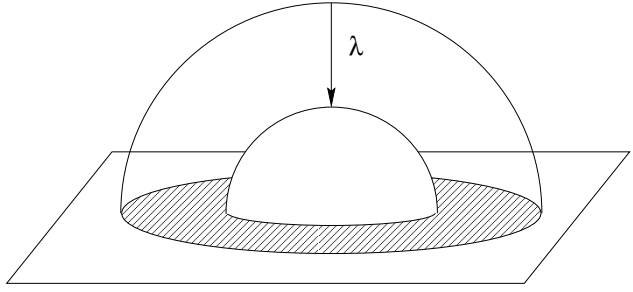,height=1.5in}}}
\caption{Genus one case.}
\label{fig:g1}
\end{figure}

The fundamental domain for a normalized Schottky group of genus one is
an annulus centered at $w=0$, see Fig. \ref{fig:g1}. The transformation
$L$ maps the upper hemisphere to the lower one. Thus, the Euclidean
space that one gets after the identifications have been made is just 
the region between the hemispheres, with the hemispheres being
identified. As is clear from the figure, this space is a
solid torus. Having put the fixed points of $L$ to $w=0,\infty$, 
the only parameter left is the multiplier $\lambda$ . The geometrical 
meaning of $\lambda$ is that its absolute value is the distance 
between the hemispheres, while its argument is the angle by which
one hemisphere is rotated with respect to the other in the 
identification. One can already recognize the Euclidean BTZ black
hole in this space, with its angular coordinate running along
the geodesics connecting the hemispheres. We discuss this relation
in more details in the next subsection, where the Lorentzian interpretation
of our spaces is considered.

The Schottky space in this case is the space of $\lambda:|\lambda|>1$.
Because of the hole $|\lambda|<1$ in it, it is not simply connected.
It is universally covered by the {\it Teichmuller} space. The Teichmuller 
space can be modeled by the upper half-plane. This is done by
introducing the modular parameter:
\begin{equation}
\tau = {{\rm Arg}\lambda\over 2\pi} + i{\ln{|\lambda|}\over 2\pi}.
\label{tau}
\end{equation}
The Schottky space is then just the strip $|{\rm Re}\tau|<1/2$. Because
of the identifications of the sides of the strip, the Schottky space
is not simply connected, and its universal cover --the Teichmuller
space-- is the whole upper half-plane. The corresponding Riemann space, 
i.e., the space of different conformal structures
of the $g=1$ case, is the quotient of this space
with respect to the action of the modular group, which in
this case is ${\rm PSL}(2,\Z)$. This is the 
general situation: the Schottky space is not simply connected, and
its universal cover is just the Teichmuller space. The Riemann 
space is a quotient of the Teichmuller (and Schottky) spaces,
see \cite{Sch-R}.

\bigskip
\noindent{\bf Example B: g=2}

This is a much more complicated case. It gives a good intuition
about what happens for higher genera. 
A $g=2$ Schottky group is generated by two elements $L_1,L_2$. Thus,
its elements are all ``words'' constructed from ``letters''
$L_1,L_2$ and their inverses. The (completion of the) set
of fixed points in this case is a Kantor set, of measure zero
in $\S$.

\begin{figure}
\centerline{\hbox{\epsfig{figure=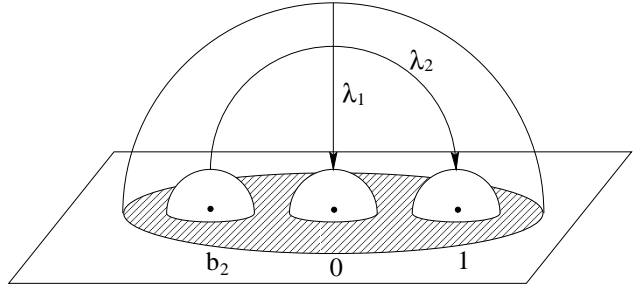,height=1.5in}}}
\caption{Genus two case.}
\label{fig:g2}
\end{figure}

A normalized Schottky group is obtained by conjugating the
generator $L_1$ so that its fixed points lie at $w=0,\infty$,
and conjugating $L_2$ to bring its attractive fixed point at
$w=1$. The fundamental region then looks like that in Fig. \ref{fig:g2}
The transformation $L_1$ identifies the upper and lower
hemispheres, while $L_2$ maps left to the right one. As is
clear from the figure, the fundamental domain becomes a genus
two surface after identifications. Thus, the Euclidean space one
gets is a solid two-handled sphere. In the next subsection we
shall see that such solid two handled spheres are Euclidean
versions of, for example, the known \cite{Brill}
single asymptotic region wormhole solution, and of three 
asymptotic region black hole.

Let us now calculate the number
of parameters describing the $g=2$ case. They are: the multipliers
$\lambda_1,\lambda_2$ of both generating transformations, plus
the complex coordinate $b_2$ of the repulsive fixed point of the
second generator. Thus, we have in total 6 real parameters.
The Schottky space, however, cannot be described so explicitly as
in the case of $g=1$. The general result only tells us that
this space is a connected region in $\C^3$. Also, particular
boundary points can be obtained by an explicit hard labor
calculation.

As in the genus one case, the Schottky space is not simply
connected; its universal cover is the Teichmuller
space of genus two. The corresponding Riemann space can be
obtained as a quotient of each of these spaces.

\subsection{Lorentzian interpretation}
\label{ss:lor}

We now proceed to the Lorentzian signature interpretation of
the spaces described above. Although the Lorentzian
spacetimes of this subsection were known before, 
the fact that these spacetimes have natural ``analytic
continuations'', where they become solid $g$-handled spheres,
to the best of our knowledge, was not realized.

We first review some known facts about the black hole solutions
in 2+1 dimensions, and then describe a procedure that, given
a black hole spacetime, constructs the corresponding Euclidean
signature space. 
 
The spacetimes we are going
to describe are constant negative curvature spaces, and
thus are obtained from Lorentzian AdS${}_3$ by discrete identifications.
The corresponding constructions are described in details 
in \cite{Brill,Rot}; we shall only sketch them here.

Let us first consider the non-rotating case \cite{Brill}. The relation
between the spaces described above and the Lorentzian signature
black holes is most transparent in this case. The case of rotating
black holes is somewhat more complicated; we discuss it at the end
of this subsection.

The characteristic feature of non-rotating spacetimes of \cite{Brill}
is that there is a surface of time symmetry. Let us call it the
$t=0$ hypersurface. Because it is time symmetric, its extrinsic
curvature vanishes. Thus, it is a two-dimensional Euclidean signature
space of constant negative curvature. Such spaces are all obtainable by 
identifications of points of AdS${}_2$, and are easy to classify
by specifying the discrete group of identifications that was used.
Having described the geometry of $t=0$ plane, one then
``evolves'' this geometry forwards and backwards in time to obtain the
whole spacetime.

\begin{figure}
\centerline{\hbox{\epsfig{figure=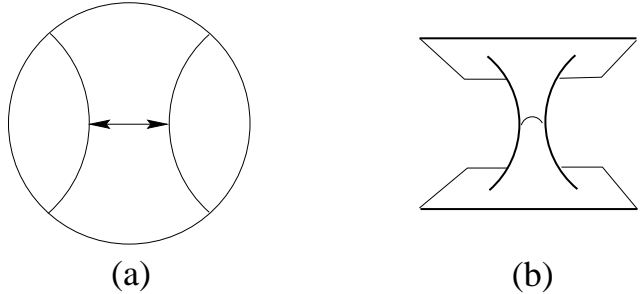,height=1.5in}}}
\caption{BTZ black hole: the initial slice geometry.}
\label{fig:btz}
\end{figure}

Let us now describe the possible $t=0$ slice geometries. There 
are two basic possibilities: the geometry can be either compact or 
non-compact. 

The compact case is well-known from the theory of uniformization
of Riemann surfaces. The discrete groups one uses are the so-called
{\it Fuchsian} groups. The fundamental region in this case is a
$4G$-sided polyhedron in AdS${}_2$. A Fuchsian group
is a discrete subgroup of the group of isometries, i.e., subgroup of $\isom$. 
It is generated by $2G$ elements,
with one relation between them. The geometries that arise are
compact genus $G$ Riemann surfaces. 

The spacetimes one gets by evolving such compact $t=0$ slice
geometries are expanding and then collapsing compact universes. In other words, they
appear at some moment in past from the past singularity, and 
disappear in the future singularity in finite time. Such spacetimes
may be interesting as simple cosmological models. They were studied
in this context in, e.g., \cite{Marolf-H}.

The black holes of \cite{Brill} realize the other possibility:
non-compact $t=0$ surface. They are still obtainable from 
AdS${}_2$ by discrete identifications by Fuchsian groups
(by which we mean discrete subgroups of $\isom$);
these are, however, not the ``classical'' Fuchsian groups,
for one gets non-compact surfaces. Thus, in this case, the initial time geometry
will have asymptotic regions. Evolving these asymptotic regions,
one gets the asymptotic regions of the corresponding
spacetime. These will generally occupy only a finite region
on the boundary of the Lorentzian AdS${}_3$. Then the boundaries
of the past of these asymptotic regions will have the interpretation
of event horizons: one gets black holes.

It is instructive to see how the usual BTZ black hole can be obtained
this way. The $t=0$ slice of the BTZ black hole is shown in 
Fig. \ref{fig:btz}(a); following \cite{Brill} we use
the unit disc model for AdS${}_2$. The corresponding discrete group 
(subgroup of ${\rm SU}(1,1)\simeq \isom$) is generated by a single element.
The fundamental region is the ``strip'' between the two
geodesics; one of them is mapped into the other by the
generator. The quotient space has the topology of the 
$S^1\times\R$ wormhole with two asymptotic regions, 
each having the topology of $S^1$, see Fig. \ref{fig:btz}(b).
The BTZ angular coordinate runs from one
geodesics to the other. The distance between the two
geodesics measured along their common normal 
is precisely the horizon circumference.

\begin{figure}

\unitlength 0.800mm
\linethickness{0.4pt}
\begin{picture}(122.87,72.61)(0,0)
\thicklines

\bezier{40}(116.60,14.31)(116.60,11.06)(108.76,9.43)
\bezier{40}(108.76,9.43)(100.93,7.81)(93.09,9.43)
\bezier{40}(93.09,9.43)(85.26,11.06)(85.26,14.31)
\bezier{40}(116.60,42.15)(116.60,39.47)(108.76,38.14)
\bezier{40}(108.76,38.14)(100.93,36.81)(93.09,38.14)
\bezier{40}(93.09,38.14)(85.26,39.47)(85.26,42.15)
\bezier{16}(105.63,37.75)(104.67,38.56)(108,40)
\bezier{22}(108,40)(111.42,40.5)(113.46,39.37)
\bezier{16}(96.23,46.54)(97.18,45.73)(94.66,44.93)
\bezier{22}(94.66,44.93)(91.44,44.23)(88.39,44.93)
\bezier{40}(116.60,69.99)(116.60,72.30)(108.76,73.45)
\bezier{40}(108.76,73.45)(100.93,74.61)(93.09,73.45)
\bezier{40}(93.09,73.45)(85.26,72.30)(85.26,69.99)
\bezier{40}(116.60,69.99)(116.60,67.68)(108.76,66.53)
\bezier{40}(108.76,66.53)(100.93,65.37)(93.09,66.53)
\bezier{40}(93.09,66.53)(85.26,67.68)(85.26,69.99)
\put(85.26,14.31){\line(0,1){55.68}}
\put(116.60,14.31){\line(0,1){55.68}}
\thicklines
\multiput(88.78,67.48)(0.50,0.12){46}{\line(1,0){0.50}}

\multiput(88.39,10.96)(0.42,0.12){58}{\line(1,0){0.42}}

\bezier{260}(88.78,67.48)(122.87,36.21)(88.39,10.96)
\bezier{74}(111.90,72.77)(117.38,64.60)(116.60,53.28)
\bezier{64}(115.03,31.57)(111.90,39.55)(115.03,47.72)
\bezier{40}(115.03,47.72)(116.60,53.28)(116.60,53.28)
\bezier{44}(115.03,31.57)(116.60,27.11)(116.60,25.44)
\bezier{22}(116.60,25.44)(116.60,20.99)(112.68,17.93)
\bezier{17}(111.90,72.77)(103.02,65.16)(98.58,56.63)
\bezier{114}(98.58,56.63)(91.53,42.89)(103.28,28.23)
\bezier{12}(103.28,28.23)(106.41,23.40)(112.68,17.93)
\bezier{74}(88.39,10.96)(84.47,19.69)(85.26,31.01)
\bezier{64}(86.82,53.84)(89.96,44.75)(86.82,36.58)
\bezier{40}(86.82,36.58)(85.26,31.01)(85.26,31.01)
\bezier{44}(86.82,53.84)(85.26,57.18)(85.26,59.97)
\bezier{22}(85.26,60.52)(85.26,63.31)(88.78,67.48)
\thinlines

\put(117.38,42.15){\makebox(0,0)[lc]{$t=0$}}
\bezier{40}(116.60,14.31)(116.60,16.25)(112.68,17.93)
\bezier{10}(112.68,17.93)(104.85,20.43)(98.58,19.87)
\bezier{8}(98.58,19.87)(90.35,19.32)(86.43,16.53)
\bezier{10}(86.43,16.53)(85.26,15.59)(85.26,14.31)
\bezier{16}(85.26,42.15)(84.94,43.71)(88.39,44.97)
\bezier{5}(88.39,44.97)(91.94,46.23)(95.91,46.60)
\bezier{40}(95.91,46.60)(99.89,46.98)(103.96,46.75)
\bezier{7}(103.96,46.75)(110.02,46.31)(114.09,44.67)
\bezier{12}(114.09,44.67)(116.18,44.00)(116.60,42.15)

\bezier{92}(89.00,67.00)(88.61,57.78)(94.00,45.00)
\bezier{140}(89.00,67.00)(103.89,56.39)(108.00,40.00)
\put(94.00,44.80){\line(3,-1){11.50}}
\put(93.00,48.00){\line(3,-1){12.00}}
\put(92.00,51.00){\line(3,-1){12.00}}
\put(91.00,54.00){\line(3,-1){11.80}}
\put(90.00,57.00){\line(3,-1){11.00}}
\put(89.50,60.00){\line(3,-1){9.00}}
\put(89.00,62.50){\line(3,-1){7.00}}

\multiput(89,65)(0.42,-0.13){8}{\line(1,0){0.42}}
\bezier{5}(105.4,41.00)(106.5,40.50)(107.99,40.00)
\bezier{3}(105.3,44.00)(106.1,43.65)(106.94,43.30)
\bezier{3}(103.94,47.00)(104.9,46.75)(105.78,46.50)
\bezier{3}(102.8,50.00)(103.65,49.75)(104.5,49.50)
\bezier{2}(101.28,53.17)(101.7,53.08)(102.11,53.00)
\bezier{2}(98.44,56.83)(99.22,56.66)(99.94,56.50)
\bezier{2}(96.44,60.00)(97.72,59.91)(97.00,59.82)

\bezier{144}(108.00,40.00)(117.78,61.11)(112.00,73.00)
\end{picture}

\caption{BTZ black hole: the spacetime picture. Two asymptotic
regions are shown: they are parts of the boundary cylinder
lying between the timelike geodesics. The future event horizon,
which is the boundary of the past of the asymptotic infinity,
is shown. Note that it intersects the initial slice along the
minimal line connecting the two geodesics bounding the fundamental
region.}
\label{fig:btz-st}
\end{figure}
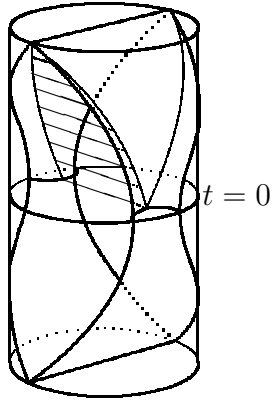

One can obtain the spacetime geometry of this black hole
by ``evolving'' the $t=0$ slice geometry. In this evolution,
geodesics of the initial slice
evolve into geodesic surfaces in the spacetime, which are then
to be identified. The region between them,
after identifications, is the BTZ black hole spacetime,
see Fig. \ref{fig:btz-st}. Because timelike geodesics in
AdS are ``attracted'' to each other, the two geodesic
surfaces will finally cross. This is where the spacetime
``ends''. Note, however, that the time it takes for them 
to meet each other is finite only in the global AdS
time coordinate. In BTZ time coordinate the corresponding
``singularities'' are at past and future infinity. The
asymptotic infinity of the BTZ black hole consists of
two regions on the AdS boundary cylinder. These are the
regions between the timelike geodesics on the boundary;
the geodesics are identified. One can now construct the
past of the asymptotic region to convince oneself that there
is a region in this spacetime that is causally disconnected
from the infinity, see Fig. \ref{fig:btz-st}. 
This region is the black (white) hole, and its
boundary is the event horizon.

\begin{figure}
\centerline{\hbox{\epsfig{figure=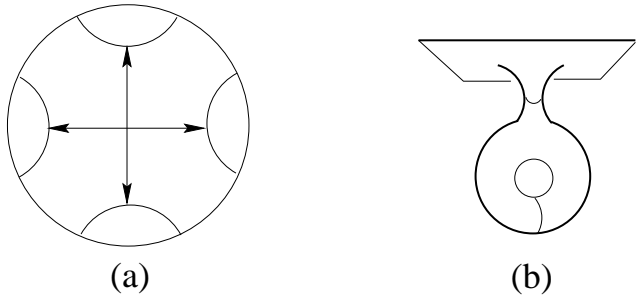,height=1.5in}}}
\bigskip
\caption{Initial slice geometry of the single 
asymptotic region black hole with a toroidal wormhole inside the horizon}
\label{fig:wormhole}
\end{figure}

Let us now, following \cite{Brill}, consider more complicated
initial slice geometries. Let the fundamental region to be
the part of the unit disc between four geodesics, as in 
Fig. \ref{fig:wormhole}(a). Let us identify these geodesics
cross-wise. It is straightforward to show that the resulting
geometry has only one asymptotic region, consisting of all
four parts of the infinity of the fundamental region. With
little more effort one can convince oneself that the resulting
geometry is one asymptotic region ``glued'' to a torus, 
see Fig. \ref{fig:wormhole}(b). Considering the spacetime
that one gets by evolving this geometry, one finds that this
is a single asymptotic region black hole, but the topology
{\it inside} the event horizon is now that of a torus.
See \cite{Brill} for more details on this spacetime.

\begin{figure}
\centerline{\hbox{\epsfig{figure=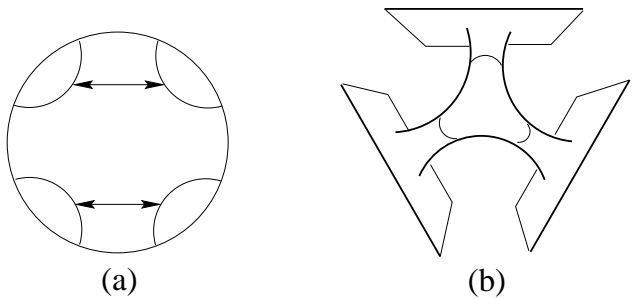,height=1.5in}}}
\caption{Initial slice geometry of the three asymptotic region black hole}
\label{fig:3bh}
\end{figure}

Another interesting and simple spacetime is that containing
three asymptotic region black hole \cite{Brill}. The fundamental
region on the $t=0$ plane is again the region bounded by four
geodesics. They are, however, now identified side-wise,
see Fig. \ref{fig:3bh}(a). One can clearly see that the
initial slice geometry has three asymptotic regions, as
in Fig. \ref{fig:3bh}(b). Evolving this, one gets a spacetime
with three asymptotic regions and corresponding event horizons.
See \cite{Brill} for more details.

This procedure allows one to construct a large variety of spacetimes,
see \cite{Brill}. In particular, one can have a single asymptotic region 
black hole with an arbitrary Riemann surface inside the horizon.

The construction of black hole spacetimes we just described 
is well-known. It was not realized, however, that there exists
a natural way to ``analytically continue'' these spacetimes
into the Euclidean signature. As we shall see in a moment,
the Euclidean spaces one obtains are just the spaces constructed in 
the previous subsection. 

All the spacetimes described were obtained
by discrete identifications of points of the Lorentzian signature AdS${}_3$.
Let us try to construct their Euclidean analogs; the basic
idea is to identify points in Euclidean AdS${}_3$ using
the ``same'' transformations as the ones that were used in the 
Lorentzian case. We must be careful, however,
about what the ``same'' means, for the groups of isometries in these
two cases are different. In the case of non-rotating black holes,
however, there is a natural choice. Indeed,
in spacetimes we described there
is a surface whose geometry locally is that of Euclidean AdS${}_2$.
Let us require that in the Euclidean spaces we are going to construct
there is also a surface of ``time symmetry'', whose geometry coincides
with the geometry of $t=0$ slice of the Lorentzian case. One can now
``evolve'' this geometry in the Euclidean ``time'' to obtain the
whole Euclidean space. It is easy to see that this procedure 
determines the Euclidean geometry completely, once the geometry
of the $t=0$ slice of the Lorentzian case is specified.

\begin{figure}
\centerline{\hbox{\epsfig{figure=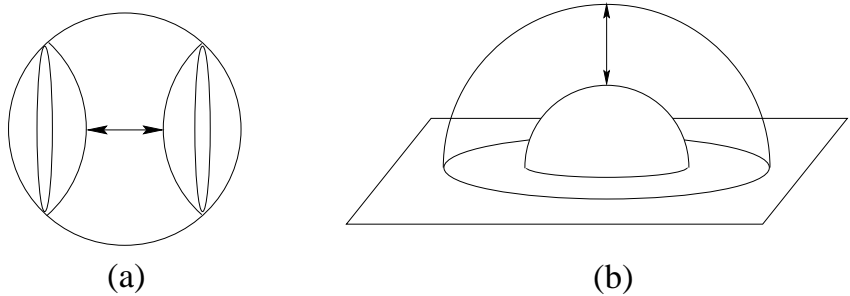,height=1.5in}}}
\caption{Euclidean BTZ black hole}
\label{fig:btz-eucl}
\end{figure}

Let us see how this works for the simplest case of BTZ black hole.
Let us use the unit ball model for the Euclidean AdS${}_3$. Thus,
we require that the geometry of one of the slices of this ball
going through the origin is the same as the geometry of the
$t=0$ slice of BTZ black hole, see Fig. \ref{fig:btz}. We then
simply have to continue the geodesics that are identified on the
$t=0$ slice to geodesics surfaces in AdS${}_3$; one gets two
hemispheres, as in Fig. \ref{fig:btz-eucl}(a). The Euclidean
BTZ black hole is then the region between these hemispheres;
the hemispheres themselves are identified. To represent this
space in the form already familiar to us, let us map the
unit ball into the upper half-space. Let us also ``normalize
things'' by putting the
fixed points to $z=0,\infty$. What we get is two hemispheres
centered at $z=0$. The Euclidean BTZ is the region between
the hemispheres. The $t=0$ slice of our original Lorentzian
signature black hole is now simply the $y=0$ slice. The 
horizon is represented in the Euclidean picture by the part
of the $x=y=0$ line that is contained between the hemispheres.
The distance between the hemispheres along that line is 
the horizon circumference. The geometry we got is, of course,
just the usual Euclidean BTZ hole geometry that can be obtained
from the Lorentzian signature BTZ metric by the analytic continuation
of the time coordinate, see \cite{Carlip-T}. But what we obtained
is also just one of the cases considered in the previous subsection.
Indeed, it is just the genus one case for the special case when
the multiplier $\lambda$ is purely real. The ``reality'' of
$\lambda$ is, of course, the consequence of the fact that we are dealing with
the non-rotating BTZ black hole. As we shall see below, allowing
$\lambda$ to be complex indeed corresponds to the inclusion
of rotation. Thus, using our ``analytic continuation'' procedure
we arrived at the known Euclidean BTZ geometry; we also see that
it is just one of the cases considered in the previous subsection,
namely, the Euclidean BTZ black hole is a solid torus.

\begin{figure}
\centerline{\hbox{\epsfig{figure=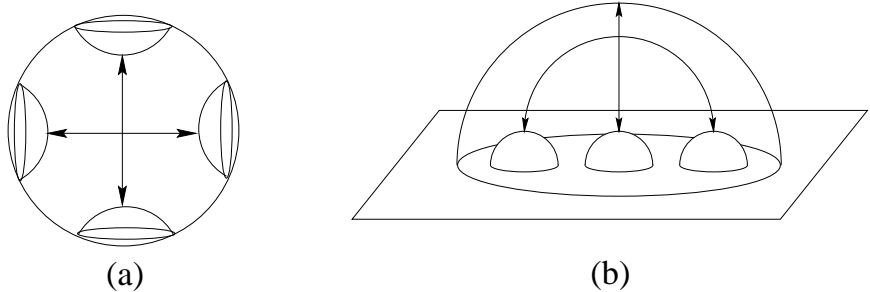,height=1.5in}}}
\caption{Euclidean single asymptotic region black hole with a torus inside}
\label{fig:wormhole-eucl}
\end{figure}

Let us consider another example. We now want to construct the
Euclidean version of the single asymptotic region black hole
with a torus inside the horizon. The procedure is the same:
we require a slice of the unit ball to have the same geometry
as the $t=0$ slice of this black hole. This gives us four
hemispheres inside the unit ball; the Euclidean space is the
region between them and they are to be identified cross-wise,
see Fig. \ref{fig:wormhole-eucl}(a).
To help the situation look more familiar, let us again map 
it into the upper half-space; we put three of the four
fixed points to $z=0,\infty,1$, see Fig. \ref{fig:wormhole-eucl}(b). 
What we get is one of our genus two case spaces described in 
the previous subsection. All three parameters $\lambda_1,\lambda_2,b_2$
are now real; this corresponds to the absence of
rotation.

One can do a similar analysis for the three asymptotic region
black hole (one also gets a solid two-handled sphere), 
and for any other of the non-rotating black holes
of \cite{Brill}. In all cases the pattern is the same: one
requires the $t=0$ slice geometry to be the same also in the
Euclidean case, and this determined the Euclidean geometry
completely. The Euclidean spaces one gets are solid $g$-handled
spheres of the previous subsection, with all parameters of the
Schottky space being real, which corresponds to no rotation.

There is a very simple formula\footnote{%
I am grateful to D.\ Brill for pointing out this formula to me.} 
that allows one to calculate
the genus of the Euclidean boundary by knowing the number of
asymptotic regions and the topology inside the horizon. Let us 
denote the number of asymptotic regions by $K$, and the genus
of the wormhole sitting inside the horizon by $G$. Then the genus
$g$ of the Euclidean boundary is given by:
\begin{equation}
g = 2G + K - 1.
\end{equation}
One can easily convince oneself that this formula holds by checking it
on few examples. To see this formula in use, let us determine the spaces that
have $g=2$. One can easily see that they are: (i) $G=1,K=1$ - 
single asymptotic region, torus inside the horizon, and (ii)
$G=0,K=3$ - three asymptotic regions.

Before we turn to the rotating case, an important remark is in order.
As is discussed in details in \cite{Brill}, the geometry of each
of the asymptotic regions of the black hole spacetimes described
is {\it always} the same as that of the asymptotic region of the
BTZ black hole of certain mass. What is different in these
black holes is only the geometry inside the horizon.
This faces us with a puzzle: on one hand, our experience with 
analytic continuations tells us that the analytic continuation
procedure is only sensitive as to what happens {\it outside}
of the horizon. Thus, one would expect that, since the geometry
outside is in all cases the same, the result of the analytic 
continuation should also be the same. On the other hand,
the procedure just described clearly gives different Euclidean
spaces, for example for the BTZ black hole and for the
single asymptotic region BH with a torus inside. Thus, the
``analytic continuation'' procedure that we described is
{\it different} from the usual procedure.

We would now like to convince a skeptical reader that this 
alternative ``analytic continuation'' procedure is more 
natural in our context than the usual one. An important observation 
here is that the spacetimes we are working with do not have a 
global Killing vector field generating time 
translation (actually, no global KVF's at all). There
only is a timelike KVF outside of the horizons, in the asymptotic 
regions. Inside the horizons, the spacetimes are non-stationary;
as discussed in \cite{Brill}, the wormhole sitting inside the
horizons evolves, and, in particular, changes its size. The
absence of a global KVF is a generic feature of all these
spacetimes (except for the BTZ case). Thus, the usual procedure
of analytically continuing in the coordinate defined by the
time translations KVF seems unnatural in our case, for there is no 
such global KVF. The procedure described above, however,
does not need any such global KVF. 
Moreover, as we saw above, it gives {\it the same} result as
the usual analytic continuation procedure in the case of the 
BTZ black hole, which does have a global time KVF. Thus, in our opinion,
the procedure described is an alternative to the usual
analytic continuation procedure, which is more natural in the
situations where no global KVF corresponding to time translations
is present. Of course, as it stands, this alternative procedure
is only available in the case
of 2+1 gravity; because of the absence of local degrees
of freedom here the spacetimes are constructed by
discrete identifications, and this is what makes our
procedure possible.

Having this said, and hoping that this convinces the reader
that the procedure described does make sense, let us now
discuss the rotating case. As usual in general relativity,
the inclusion of rotation is much harder than one could naively
expect it to be. So far, the rotating ``wormholes'' are much less
studied than non-rotating ones. There are essentially only few
references on the subject, see \cite{Rot,Match,Brill-new}. The trick with
the time symmetry surface, that greatly simplified the classification
of the non-rotating spacetimes, is not available, because
there is now no such surface. To construct such spacetimes one
takes a different route \cite{Rot} of studying the action of the discrete
group of identifications on the AdS${}_3$ boundary cylinder.
This way one can analyze the causal structure of the 
resulting spacetimes, and it is possible to explicitly compute the
mass and angular momentum parameters of the asymptotic regions,
see \cite{Rot} for details. We are not going to use this 
construction here. Instead, we shall list some facts that, we
hope, will convince the reader that the general case Euclidean
spaces considered in the previous subsection indeed correspond
to rotating black holes. 

We have already encountered such facts before. Indeed, we saw
that the Euclidean spaces corresponding to non-rotating black holes are 
just {\it special} points in the Schottky space, in the sense that
all parameters are real. When going to the rotating case, one
naturally expects the number of parameters to be doubled.
This is, of course, exactly what happens when one allows the
parameters of the Schottky space to be complex. 
More evidence is given by the case of BTZ black hole. The
Euclidean space corresponding to this black hole is described
in \cite{Carlip-T}. The only difference between the rotating and
non-rotating cases is that in the rotating case 
the two hemispheres of Fig. \ref{fig:btz-eucl}(b)
are rotated with respect to each other when one performs the identification.
The amount of this rotation encodes information about the
hole angular momentum, see \cite{Carlip-T}. But such
identification with rotation exactly corresponds to making
the multiplier $\lambda$ of the transformation complex. Thus, in the case
of BTZ black hole, the presence of rotation indeed corresponds 
in the Euclidean case to the transformation parameter being
complex. All this strongly suggest that the inclusion of
rotation indeed corresponds to making the Schottky parameters
complex. We shall not further dwell on this issue here. 
We hope to study the details of the described  
``analytic continuation'' procedure in future publications.

To summarize, we have argued that the Euclidean spaces studied in 
the previous subsection correspond, in the Lorentzian signature,
to the black holes of complicated topology. Interestingly,
this, together with our next section results on the stress-energy tensor of
the boundary CFT, provides an approach to the problem of classification
of such spacetimes; they are now classified by their Euclidean
counterparts, which are completely under control. This finishes
our excursion to the zoo of 2+1 dimensional black holes. From now
on we concentrate on the Euclidean spaces, and consider the problem
of calculation of the partition function of the holographic CFT.

\section{Partition function}
\label{sec:part}

To obtain the semi-classical approximation to the partition function,
we first need to find all spaces that have a fixed behavior at infinity,
and calculate the on-shell classical action for these spaces. We
first consider the problem of computation of the on-shell action.
We then look, in Subsection \ref{ss:therm} at various issues 
related to the thermodynamics of our spaces. Finally, in Subsection
\ref{ss:cft}, we study the question of how to perform the sum
over {\it all} spaces having the same behavior at infinity,
and discuss the properties of the arising partition function.

\subsection{Regularization. Conformal anomaly}
\label{ss:anomaly}

Describing the idea of holography in the introduction, we
were not specific enough about what is meant 
``appropriately rescaled boundary value of the metric'',
which is to be kept fixed when one performs the bulk
path integral.
In asymptotically AdS spaces the boundary at infinity
can be brought at a finite distance by a conformal completion
of the space. In this procedure
one constructs a new, ``unphysical'' metric, conformally related
to the original ``physical'' one. The boundary is now at a finite
distance in this new metric, and one can restrict the unphysical
metric to the boundary. The boundary metric one gets is, however,
defined only up to conformal transformations,
as is clear from the construction. Thus, the only invariant
information in the boundary metric is the conformal structure
it defines. Thus, it is this conformal structure at infinity 
that one would like to keep fixed when summing over the bulk 
metrics in the gravity path integral.

However, already at the level of the semi-classical approximation,
the following problem arises. As we have said, in this approximation
one sums over all classical solutions with the prescribed conformal
structure at infinity. The contribution of each classical solution
is the exponential of the Einstein-Hilbert action evaluated on this
solution. However, because the space is non-compact, 
this generally diverges. The standard way to deal with this divergence
is to use some subtraction procedure. Thus, one first regulates the
action by integrating only over a portion of the space. Some of the
pieces of the regulated action diverge as the ``cutoff'' is removed.
To obtain a regularized action, one subtracts the divergent pieces, 
and then removes the cutoff. In the context of asymptotically flat
spacetimes it is usually enough to subtract the integral of the
extrinsic curvature of the boundary embedded in the flat space. A
natural AdS analog of this procedure would be to subtract the volume
of AdS space itself; this is what was done in the original \cite{HP}
discussion of the black hole thermodynamics in AdS. However, it turns out
that for a general asymptotically AdS space the volume subtraction
is not enough; there is an additional type of divergence that should
be dealt with: the so-called logarithmic divergence. After removing this 
divergence the regularized action generally will depend 
not only on the conformal structure of the
metric at infinity, but also on a particular representative that
was used in the regularization procedure. In other words, the
regularized action is not invariant under conformal transformations:
one gets the anomaly \cite{H-Skenderis}. The good news is that this
anomaly is mild: it is the usual conformal anomaly to which one
is accustomed in conformal field theory. This anomaly is exactly
that of a conformal field theory of central charge $c=3l/2G$, in
accordance with the old result \cite{Brown-H}. 

In this subsection we calculate the conformal anomaly for our
spaces. For metrics of
constant curvature, the on-shell Einstein-Hilbert action is 
proportional to the volume of the space:
\begin{equation}
{1\over 16\pi G} \int d^3x \, \sqrt{g} (R-2\Lambda) =
{\Lambda\over 4\pi G} \int d^3x \, \sqrt{g} = - {V\over 4\pi G l^2},
\end{equation}
where we have introduced the notation $V$ for the volume of space,
and the cosmological constant $\Lambda=-1/l^2$. 

Let us now consider the volume of the spaces of the previous section. 
The simplest regularization is to introduce a plane at $z=\epsilon$
and consider the volume above this plane. The calculation is then
straightforward. One has to find the volume contained inside the
``big'' hemisphere built on the circle $C_1$ and subtract from it the
volumes inside all other hemispheres. The hyperbolic
volume contained inside a hemisphere of radius $\rho$ and above the 
$z=\epsilon$ plane is easy to calculate. The result is:
\begin{equation}
V_\epsilon(\rho)=\pi l^3 \left( 
{\rho^2\over 2\epsilon^2} - {1\over 2} - \ln{{\rho\over\epsilon}}
\right).
\end{equation}
We thus see, that, as the cutoff is removed, there are two 
types of divergences: $1/\epsilon^2$ and $\ln{\epsilon}$. 
As is easy to see, the first divergence is the usual ``area of the
boundary'' type divergence, present for any asymptotically
AdS space. Indeed, the volume $V$ of any asymptotically
AdS space grows as $Al/2$, where $A$ is the area of the boundary.
This divergence can be removed by adding a boundary term, local
in the boundary metric, to the bulk action \cite{H-Skenderis,Vijay}. 
This boundary term is a multiple of the boundary area. The second,
logarithmic divergence can also be removed by adding a local
counterterm. This, however, spoils the conformal invariance of
the regularized action, as we discussed before. 

Thus, the regularized action can be written as:
\begin{eqnarray} \nonumber
I[\gamma]=&{}& -{1\over 16\pi G} \left(
\int d^3x \, \sqrt{g} (R - 2\Lambda) + 2 \int d^2x \, \sqrt{\gamma} \, K -
{2\over l} \int d^2x \, \sqrt{\gamma} \right) + \\
&{}& \left(\lower1ex\vbox{\hbox{{\rm \quad counterterm\,\,removing}}
\hbox{{\rm the\,\,logarithmic\,\,divergence}}} \right).
\label{action}
\end{eqnarray}
Here $\gamma$ is the metric induced on the boundary, and $K$ is the
trace of the extrinsic curvature of the boundary. When computing
$I$, one keeps the boundary at a finite distance, computes all
the terms, and then removes the ``cutoff'' by sending the
boundary at infinity. The minus sign in front of the Einstein-Hilbert action 
is introduced so that the (unregularized) on-shell action is 
positive-definite. With this choice for the sign, the
regularized action is just the free energy of the corresponding
space. The regularization procedure (\ref{action}) is
the same as that of Refs. \cite{H-Skenderis,Vijay}. 

To calculate the regularized action we note that, 
in the asymptotically AdS case, the trace
of extrinsic curvature is in the leading order proportional to the
area form of the boundary, and its subleading term vanishes too
rapidly to give any contribution to the free energy $I$. Thus, 
when calculating the free energy, the integral of $K$ can be
replaced by a multiple of the total area of the boundary. Doing
this one finds, that the regularized action can be written as:
\begin{equation}
I = {V\over 4\pi G l^2} - {A\over 8\pi G l} + ({\rm counterterm}).
\end{equation}
For our spaces the volume is given
by the $\epsilon$-plane regularized volume inside the ``big'' hemisphere
minus the regularized volumes inside all other hemispheres. 
Noting that the area inside the circle, along which a 
hemisphere of radius $\rho$ intersects the $z=\epsilon$ plane is:
\begin{equation}
A_\epsilon(\rho) = {\pi l^2\over\epsilon^2} (\rho^2 - \epsilon^2),
\end{equation}
we get:
\begin{equation}
{V_\epsilon\over 4\pi G l^2} - {A_\epsilon\over 8\pi G l}
= {l\over 4G} \left( - \ln{\rho_1\over\rho_1'} + 
\ln{\rho_2\rho_2'\cdots\rho_g\rho_g'} -
(2g-2) \ln{\epsilon} \right).
\label{action1}
\end{equation}
To get the regularized action, we have to remove the logarithmically
divergent term. Let us now show that this term is just  
correct to give the conformal anomaly expected for a genus $g$
surface. The infinitesimal form of the conformal transformation 
$\gamma\to e^{2\phi}\gamma$ is 
$\delta\gamma=2\delta\phi\,\gamma$.
Then $\delta\ln{\epsilon}=-\delta\phi$. Thus, after the logarithmically
divergent term is removed, the change of the regularized action
under an infinitesimal conformal transformation is:
\begin{equation}
\delta I = (-)(- {l\over 4G}\, (2g-2))(-\delta\phi).
\label{anomaly-1}
\end{equation}
Here the first minus sign comes about because the change in the
regularized action is minus the change in the term that is removed.
The above expression must be compared with the usual formula for
the conformal anomaly:
\begin{equation}
\delta I = {c\over 24\pi} \int \sqrt{\gamma} R \,\delta\phi.
\label{anomaly-2}
\end{equation}
Recalling that we consider constant conformal rescalings, and that
for a genus $g$ surface
\begin{equation}
\int \sqrt{\gamma} R = 4\pi (2-2g),
\end{equation}
we see that the two expressions (\ref{anomaly-1}),(\ref{anomaly-2})
for the anomaly agree if the value of the central charge is
$c=3l/2G$, as expected. Thus, after removing the logarithmically
divergent term, we get the conformal anomaly expected for
a genus $g$ surface.

Removing the anomaly, we get the regularized action:
\begin{equation}
I = {l\over 4G} \left( 
- \ln{\rho_1\over \rho_1'} + \ln{\rho_2\rho_2'\cdots\rho_g\rho_g'} \right) \,.
\label{action-reg}
\end{equation}
The interpretation of this expression
is as follows. As we discussed above, the regularized 
action is anomalous, in the sense that it depends not only on 
a conformal class of the metric on $X$, but also on a particular 
representative chosen. The expression (\ref{action-reg}) is
the action calculated on the flat metric of the $\epsilon$-plane.
What we are interested in, however, is the regularized action computed
on the canonical (constant negative curvature) metric on $X$. One
could, in principle, find the conformal factor that maps the 
flat metric of the $\epsilon$-plane into the canonical genus $g$ surface
metric, and then add to (\ref{action-reg})
a multiple of the Liouville action evaluated on this conformal factor.
This would give the desired regularized action on the canonical
metric. We, however, choose another method of deriving this action.
We shall modify the regularization procedure so that the metric
induced on the regularizing surface goes to the canonical metric on
$X$ as one removes the regulator. To see, how this works, let us
first consider the $g=1$ case.

\subsection{Genus one case}
\label{ss:genus-one}

\begin{figure}
\centerline{\hbox{\epsfig{figure=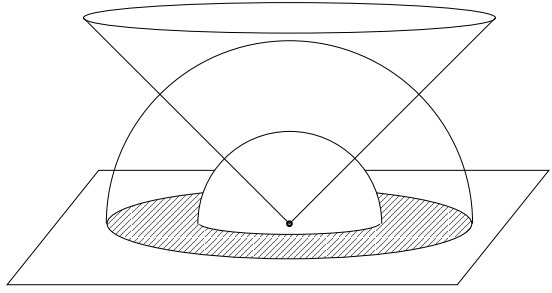,height=1.5in}}}
\caption{The ``canonical'' regularizing surface for the genus one case is a cone.}
\label{fig:cone}
\end{figure}

In the genus one case the space is the region between two hemispheres,
the hemispheres being identified. In this case the desired family of regularizing
surfaces consists of cones with the tip placed at $w=0$, see
Fig. \ref{fig:cone}. As the angle $2\alpha$ at the tip of the
cone increases $\alpha\to\pi/2$, the surface 
approaches the boundary. The metric induced on the surface of the cone is:
\begin{equation}
ds^2 = {l^2\over r^2 \cos^2\alpha} 
\left( dr^2 + r^2 \sin^2\alpha d\phi^2 \right),
\end{equation}
and as $\alpha\to\pi/2$ this behaves as a diverging factor 
$(1/\cos^2\alpha)$ times the canonical torus metric, 
the torus coordinates being $\ln r/2\pi, \phi/2\pi$. 
Thus, this family of surfaces has the desired property that the
induced metric approaches in the limit the canonical one.
Note that this family of regularizing surfaces is the same as the one used
in calculations of the black hole partition function \cite{HP,Witten-AdS}.
Indeed, the cones are just the spheres $r=const$ in the usual 
Schwarzschild-like coordinates. Let us now compute the regularized action.
The volume of the region between the two hemispheres and inside the cone
is given by:
\begin{equation}
V_\alpha = \pi l^3 \left( \ln{\rho\over\rho'}\right)
\left({1\over \cos^2\alpha} - 1 \right).
\end{equation}
The area of the part of the surface of the cone contained 
between the hemispheres is:
\begin{equation}
A_\alpha = 2\pi l^2 \left( \ln{\rho\over\rho'}\right) \,
{\sin\alpha\over \cos^2\alpha}.
\end{equation}
Thus, the regularized action is:
\begin{equation}
{V_\alpha\over 4\pi G l^2} - {A_\alpha\over 8\pi G l} =
{l\over 4G}\,\left( \ln{\rho\over\rho'}\right) \,
{1-\cos^2\alpha-\sin\alpha\over \cos^2\alpha},
\end{equation}
which, in the limit $\alpha\to\pi/2$ gives
\begin{equation}
I = - {l\over 8G}\,\ln{\rho\over\rho'} = 
- {l\over 8G}\,\ln|\lambda| \,,
\label{I-one}
\end{equation}
where $\lambda$ is the multiplier of the transformation used to 
identify the hemispheres.
This is the regularized action for the canonical metric on the
torus. It is not hard to see that the result (\ref{I-one}) reproduces correctly
the free energy of AdS and BTZ spacetimes. For the case of periodically 
identified in the imaginary time AdS, $\ln|\lambda|$ is just the period $\beta$
of the imaginary time. Thus, for AdS:
\begin{equation}
I = -{\beta\over 8G}, \qquad M = {\partial I\over\partial\beta} =
-{1\over 8G},
\end{equation}
and the entropy is zero because the free energy depends linearly on $\beta$.
In the case of the BTZ spacetime, $\ln|\lambda|$ is the horizon
circumference $2\pi r_+$, and $\beta= 2\pi l^2/r_+$. Thus,
\begin{equation}
I = - {1\over 8G} {(2\pi l)^2\over\beta}, \qquad \qquad M = 
{\partial I\over\partial\beta} = {1\over 8G} {(2\pi l)^2\over\beta^2} =
{1\over 8G} {r_+^2\over l^2},
\end{equation}
and
\begin{equation}
S = \beta M - I = {1\over 4G} {(2\pi l)^2\over\beta} = {2\pi r_+\over 4G},
\end{equation}
which are the usual results for BTZ black hole.

We would now like to generalize the above calculation to the
case of an arbitrary genus. The key problem here is to find
a family of regularizing surfaces such that the induced metric
approaches the canonical metric as the regulator is removed.
This can be done using the solution to the Liouville equation, as
we will show in \ref{ss:reg}. However, before we proceed with
this construction, we need to review some relevant facts
on Liouville theory.

\subsection{Uniformization. Liouville theory. Takhtajan-Zograf action.}
\label{ss:unif}

Here we recall some facts about uniformization of Riemann surfaces by
Kleinian groups, and the relation of this to Liouville
theory. We shall also introduce the Takhtajan-Zograf action.
The material reviewed here is from work \cite{Takht}.

One arrives at the Liouville equation considering the following
problem. For each of the Riemann surfaces arising as the boundary of our
spaces, we need to find the canonical curvature $R=-1$ metric that is in the same 
conformal class as the one induced by the flat metric. This is done
as follows. Let us introduce complex coordinates $w$ on $\S$. The
flat metric on $\S$ is $ds^2 = |dw|^2$. The constant curvature
metric we are looking for is related to the flat metric by a
conformal transformation:
\begin{equation}
d\tilde{s}^2 = e^{2\phi(w,\bar{w})}\,ds^2.
\end{equation}
The field $\phi(w,\bar{w})$ is a real field on $\S$. The condition
that $d\tilde{s}^2$ has constant curvature $R=-1$ then translates into an
equation on $\phi$:
\begin{equation}
\Delta \phi = 4 \partial_w \partial_{\bar{w}} \phi = {1\over 2} e^{2\phi}.
\label{liouv-eqn}
\end{equation}
This is the famous Liouville equation. It is rather hard to
solve in practice, except for the genus one case, but an indirect
way of describing the solution is as follows. Let us recall that
any Riemann surface $g>1$ can be uniformized by the upper half-plane $H$.
This means, that, given a Riemann surface $X$, there is a projection
map $\pi_\Gamma:H\to X$, which is just the quotient map of $H$ by
the action of some (Fuchsian) discrete group $\Gamma$ of isomorphisms 
of $H$. Moreover, the group $\Gamma$ is isomorphic to the
fundamental group $\pi_1(X)$. The Riemann surfaces $X$ arising in our
construction are quotients of the complex plane $\S$. Let us uniformize them
by the hyperbolic plane. We get a commutative diagram:
\begin{equation}
\begin{diagram}
\node{H} \arrow[2]{e,t}{J} \arrow{se,b}{\pi_\Gamma}
\node[2]{\Omega} \arrow{sw,r}{\pi_\Sigma} \\
\node[2]{X}
\end{diagram}
\end{equation}
Here $\pi_\Sigma$ is the quotient map $\pi_\Sigma:\Omega\to\Omega/\Sigma=X$,
where $\Sigma$ is the corresponding Schottky group. We are interested
in the properties of the covering map $J:H\to\Omega$. The group
of automorphisms of this map is isomorphic to the subgroup $N\subset\Gamma$,
which is the smallest normal subgroup containing the elements $A_i\in\Gamma$.
This map can be considered as a function on $H$, automorphic with
respect to $A_i$, that is, $J\circ A_i=J$, and such that 
$J\circ B_i = L_i \circ J$. The inverse map $J^{-1}$ can be 
considered as a multi-valued function on $\S$. Let us denote by $z$
the complex coordinate on the hyperbolic plane $H$. Then the 
constant curvature $R=-1$ metric on $H$ is $|dz|^2/({\rm Im}z)^2$.
We thus see, that, given the inverse map $J^{-1}$, the conformal
factor mapping the flat metric $|dw|^2$ on $\S$ to the constant
negative curvature metric on $H$ is:
\begin{equation}
e^{2\phi(w,\bar{w})} = {|(J^{-1})'(w)|^2\over ({\rm Im} J^{-1}(w))^2},
\label{phi}
\end{equation}
where prime denotes the derivative with respect to $w$.  
It does not matter which branch of the function $J^{-1}$ is chosen.
Thus, having constructed the uniformizing map $J$, we have
the solution $\phi$ to the Liouville equation (\ref{liouv-eqn}).
From the above representation for $\phi$, it follows that
it has the following transformation property:
\begin{equation}
\phi(Lw) = \phi(w) - {1\over 2}\, \ln{|L'(w)|^2}.
\label{trans-law}
\end{equation}
It can be shown that there is a unique solution to the Liouville
equation (\ref{liouv-eqn}) on $\S$ with this transformation property.
We also note, for further purposes, that the Liouville stress-energy
tensor for $\phi$ is equal to the Schwarzian derivative of $J^{-1}$:
\begin{equation}
T_w = 2\phi_{ww} - 2\phi_w^2 = \{ J^{-1}(w);w \},
\label{energy1}
\end{equation}
where $\{ \cdot\, ; \cdot \}$ is the Schwarzian derivative.
This solution to the Liouville equation will be used in the next
subsection to construct a ``canonical'' family of regularizing
surfaces.

We will also need the action of Liouville theory on Riemann
surfaces defined by Takhtajan and Zograf \cite{Takht}. As before,
a Riemann surface of genus $g$ is thought of as the quotient of 
the complex plane by the action of a Schottky group of genus $g$.
The action of Takhtajan and Zograf is the following functional of the
Liouville field:
\begin{eqnarray} \nonumber
I[\phi] = - {l\over 16\pi G} \left[
\int\!\!\!\!\int_D {i\over 2} \, dw\wedge d\bar{w} \left(
4\partial_\omega \phi \partial_{\bar{w}} \phi + {1\over 2} e^{2\phi} 
\right) +  \right.  \\ \left. 
i \sum_{i=2}^g \int_{C_i} \left( d\bar{w} \,\phi {\bar{L_i''}\over\bar{L_i'}} -
dw \,\phi {L_i''\over L_i'} \right) -
{1\over 2i} \sum_{i=2}^g \int_{C_i} \ln{|L_i'|^2} \, {L_i''\over L_i'} dw
+ 4\pi \sum_{i=2}^g \ln{|l_i|^2} \right ] .
\label{action-takht}
\end{eqnarray}
Here the first integral is taken over the fundamental region $D\subset\S$,
which, in the case when the Schottky group is normalized, is the disc 
bounded by the circle $C_1$ minus the discs bounded by all other circles.
The generators $L_i$ are thought of here as functions $L_i w$ on the complex plane. 
They can be represented in the normal form as:
\begin{equation}
{L_i w - b_i\over L_i w - a_i} = \lambda_i {w - b_i\over w - a_i}, 
\qquad |\lambda_i| > 1.
\end{equation}
Here $a_i, b_i$ are the attracting and repelling fixed points of $L_i$,
and $\lambda_i$ is the multiplier of the transformation. The quantity
$l_i$ in the last term in (\ref{action-takht}) is the lower left corner matrix 
element of the matrix representation of $L_i$:
\begin{equation}
l_i = {(\lambda_i-1)\over (a_i-b_i)\sqrt{\lambda_i}}.
\end{equation}
We use the anti-clockwise orientation of contours $C_i, i>1$.

The action functional (\ref{action-takht}) has a number of important
properties. First, the equation of motion that follows from it
is just the Liouville equation. In fact, the first set of boundary 
terms in (\ref{action-takht}) is precisely such that its variation
cancels the boundary terms arising in the variation of the ``bulk''
term, so that the variational principle is well-defined. Second, as
one can explicitly check, by changing the position
of one of the pair of circles $C,C'$, the action (\ref{action-takht}) is 
independent of a choice of the fundamental region that was 
used to evaluate it. The second set of boundary terms is necessary 
for this property to hold. The last set of constants is precisely such that,
when the action is evaluated on the solution to the Liouville equation,
its variation with respect to the moduli gives the correct 
(Liouville) stress-energy tensor, see \ref{ss:energy} below.
As we will demonstrate in the next subsection, the action
(\ref{action-takht}) evaluated on the solution to the Liouville
equation coincides with the regularized volume of the corresponding 
Euclidean space.

\subsection{``Canonical'' regularization procedure}
\label{ss:reg}

\begin{figure}
\centerline{\hbox{\epsfig{figure=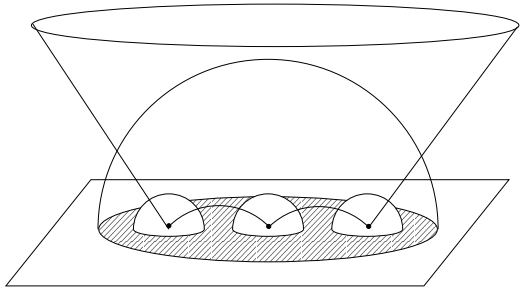,height=1.5in}}}
\caption{The ``canonical'' regularizing surface is constructed using
the Liouville field.}
\label{fig:reg}
\end{figure}

In order to find the regularized action on the canonical metric, 
we need to find a family of regularizing
surfaces such that the induced metric approaches the canonical one
when the regulator is removed. To construct a desired family of 
regularizing surfaces we use the solution of the Liouville equation. 
To this end we use the Liouville field as the coordinate $z$ running 
orthogonally to the boundary, as suggested by the work of Polyakov 
\cite{Polyakov} on non-critical string theory.
More precisely, let us introduce a family of surfaces:
\begin{equation}
z(x,y) = \epsilon e^{-\phi(x,y)}.
\label{surf}
\end{equation}
The induced metric is then given by:
\begin{equation}
ds^2 = {l^2\over \epsilon^2} e^{2\phi} \left(
dx^2 + dy^2 + \epsilon^2 e^{-2\phi} \left({\partial\phi\over \partial x} dx+
{\partial\phi\over \partial y} dy \right)^2 \right).
\end{equation}
As one can easily see, the induced metric behaves,
when $\epsilon\to 0$, as $1/\epsilon^2$ times the canonical metric on 
the Riemann surface. A typical surface looks like the one shown
in Fig. \ref{fig:reg}. It touches the boundary at points where the Liouville
field diverges. 

It is not hard to see that the prescription (\ref{surf}) 
generalizes the family of surfaces used in the genus one case. Indeed,
in this case, the Liouville field can be found introducing the
polar coordinates on $\S$ centered at $w=0$. Then the flat
metric on the annulus is:
\begin{equation}
ds^2 = |dw|^2 = dr^2 + r^2 d\theta^2 = e^{2\ln{2\pi r}}\left( 
\left ({d\ln{r}\over 2\pi} \right)^2 + 
\left({d\theta\over 2\pi}\right)^2 \right).
\end{equation}
Thus, after identifications are made, $\ln{r}/2\pi$ and $\theta/2\pi$ are
coordinates on the torus, and $\phi=-\ln{2\pi r}=-\ln|2\pi w|$ is the Liouville
field. We see that the family of surfaces (\ref{surf}) in this case
is just the family of cones considered in \ref{ss:genus-one}.

Let us now use the regularizing surfaces (\ref{surf}) to compute the
action. We would like to show that the regularized action, 
that is, the regularized volume contained in the region between the hemispheres 
and above the regularizing surface, in the limit when the regulator is removed, 
goes to the Liouville theory action (\ref{action-takht}) evaluated on the
solution to the Liouville equation. 

Before we go to the higher genus case, let us see how the action 
(\ref{action-takht}) reproduces the 
partition function for $g=1$. 
We notice that there are only contributions from the first 
term in (\ref{action-takht}). Indeed, in the genus one case,
the term involving $e^{2\phi}$ is simply not there, because the 
Liouville equation in this case is just $\Delta \phi =0$. All other
terms are zero because $L''=l=0$. Thus, evaluating
the action on the Liouville field $\phi = -\ln 2\pi r$, we get:
\begin{equation}
I = -
{l\over 16\pi G} \left [
\int_0^{2\pi} d\theta \int_{\rho'}^{\rho} rdr (1/r)^2 \right]
= - {l\over 8G} \ln|\lambda|, 
\end{equation}
which coincides with the expression for the partition function 
obtained before, see (\ref{I-one}).

Let us now do the general case. The volume is given by:
\begin{equation}
V_\epsilon = l^3 \int\!\!\!\!\int_{D_1^\epsilon} dx dy 
\int_{z_{\rm min}(x,y)}^{z_{\rm max}(x,y)} {dz \over z^3} \,,
\end{equation}
where the $x, y$ integral is taken over the region 
$D_1^\epsilon$ inside the curve $C_1^\epsilon$, which 
is the projection on the $z=0$ plane of the curve along
which the regularizing $\epsilon$-surface intersects the
hemisphere built on the circle $C_1$. Thus, the points
lying on $C_1^\epsilon$ satisfy the equation $z^2+|w|^2=\rho_1^2$.
The upper limit of the integral over $z$ is:
\begin{equation}
z_{\rm max}(x,y)^2 = \rho_1^2 - (x^2+y^2).
\end{equation}
The lower limit is different in different regions. We decompose
\begin{equation} 
D_1^\epsilon = D^\epsilon \cup {D_1^\epsilon}' 
\cup_{i=2}^g \left( D_i^\epsilon \cup {D_i^\epsilon}' \right).
\end{equation}
Within the region $D^\epsilon$ the lower integration limit is given
by the regularizing surface. In all other regions it is given by
the hemispheres built on $C_1', C_i, C_i', i>1$. Thus, we have:
\begin{eqnarray}\nonumber
V_\epsilon = l^3 \int\!\!\!\!\int_{D^\epsilon} dx dy {1\over 2\epsilon^2} e^{2\phi}
- l^3 \int\!\!\!\!\int_{D_1^\epsilon} {dx dy\over 2(\rho_1^2 - (x^2+y^2))} 
+ l^3 \int\!\!\!\!\int_{{D_1^\epsilon}'} {dx dy\over 2((\rho_1')^2 - (x^2+y^2))} \\
\label{volume}
+ l^3 \sum_{i=2}^g 
\int\!\!\!\!\int_{D_i^\epsilon} {dx dy\over 2(\rho_i^2 - ((x-x_i^0)^2+(y-y_i^0)^2))} 
\\ \nonumber
+ l^3 \sum_{i=2}^g 
\int\!\!\!\!\int_{{D_i^\epsilon}'} 
{dx dy\over 2((\rho_i')^2 - ((x-{x_i^0}')^2+(y-{y_i^0}')^2))}\,. 
\end{eqnarray}
Here $x_i^0,y_i^0$ (and the corresponding primed quantities)
are the coordinates of the center of the corresponding hemisphere.
On the other hand, the area computed using the induced metric is equal to:
\begin{equation}
A_\epsilon = l^2 \int\!\!\!\!\int_{D^\epsilon} dx dy {1\over\epsilon^2}
e^{2\phi} \left( 1 + \epsilon^2 e^{-2\phi} \left(
\left({\partial\phi\over \partial x} \right)^2 +
\left({\partial\phi\over \partial y} \right)^2 \right) \right)^{1/2},
\end{equation}
which for $\epsilon\to 0$ behaves as:
\begin{equation}
A_\epsilon = l^2 \int\!\!\!\!\int_{D^\epsilon} dx dy {1\over\epsilon^2} e^{2\phi} 
+ {l^2\over 2} \int\!\!\!\!\int_{D^\epsilon} dx dy \left(
\left({\partial\phi\over \partial x} \right)^2 +
\left({\partial\phi\over \partial y} \right)^2 \right).
\end{equation}
The first term here will cancel the first term on the right hand side
of (\ref{volume}). The second term is just the kinetic term of the
Liouville action (\ref{action-takht}). Let us compute the remaining
terms on the right hand side of (\ref{volume}). They can be reduced to 
contour integrals by introducing the polar coordinates. The typical 
integral becomes:
\begin{equation}
\int_0^{2\pi} d\varphi \int_0^{r(\varphi)} {r dr \over 2(\rho_i^2 - r^2)} =
{1\over 4} \int_0^{2\pi} d\varphi \ln{\rho_i^2\over \rho_i^2 - r^2(\varphi)}\,.
\end{equation}
This can be further transformed by noticing that on the curve
$C^\epsilon$ we have $\rho_i^2-r^2(\varphi) = \epsilon^2 e^{-2\phi}$.
Thus, the above integral reduces to:
\begin{equation}
l^3 \pi \ln{\rho_i\over\epsilon} + {l^3 \over 2} 
\int_{C_i^\epsilon} \phi,
\end{equation}
where the last integral is the contour integral of the Liouville
field. We should now remove the $\ln\epsilon$ divergence, 
combine all the pieces, and then remove the regulator. 
We get the correct ``bulk'' part of (\ref{action-takht}) plus
a set of other terms:
\begin{eqnarray}\label{*3}
&{}&I = ({\rm ``bulk''\,\,part}) \\ \nonumber
&{}& - {l\over 16\pi G} \left( 4\pi \ln{\rho_1\over\rho_1'} - 
4\pi\ln{\rho_2\rho_2'\cdots\rho_g\rho_g'} + 2\int_{C_1} \phi -
2\int_{C_1'} \phi - 2\sum_{i=2}^g \int_{C_i} \phi -
2\sum_{i=2}^g \int_{C_i'} \phi \right).
\end{eqnarray}
This can be further simplified by using the transformation
property (\ref{trans-law}) of $\phi$. The difference of the
integrals over $C_1, C_1'$ then cancels the term proportional
to $\ln{\rho_1/\rho_1'}$, and we are left with:
\begin{equation}
I = ({\rm ``bulk''\,\,part}) - {l\over 16\pi G} \left(
- 4\pi\ln{\rho_2\rho_2'\cdots\rho_g\rho_g'}
- 4\sum_{i=2}^g \int_{C_i} \phi 
+ \sum_{i=2}^g \int_{C_i} \ln{|L_i'|^2} \,\right).
\label{*1}
\end{equation}
The radii $\rho_i'$ of the circles $C_i'$ can be expressed
through the radii $\rho_i$ and the parameters of the 
corresponding Schottky group generators. The radius
$\rho_i'$ of the circle $C_i'=L_i(C_i)$ is given by:
\begin{equation}
\rho_i'={\rho_i\over |l_i|^2 (\rho_i^2-|\xi_i-w_i^0|^2)},
\end{equation}
where $w_i^0$ is the center of the circle $C_i$, and 
$\xi_i$ is the pole of $L_i$ thought of as a function on the
complex plane:
\begin{equation}
\xi_i = {\lambda_i b_i - a_i\over \lambda_i - 1}.
\end{equation}

In order to compare (\ref{*1}) to the terms in the Takhtajan-Zograf
action we notice that, the regularizing surfaces being chosen as
in (\ref{surf}), the position and radii of the circles $C_i, C_i'$
cannot be chosen arbitrarily. Indeed, one has to make sure that, in
the limit $\epsilon\to 0$, the circles $C_i^\epsilon$ are mapped
into the circles ${C_i^\epsilon}'$. As is not hard to see, the
condition for this, for each pair of circles, is:
\begin{equation}
{(\rho_i')^2 - |L_iw-{w_i^0}'|^2\over |L_i'|^2} = \rho_i^2 - |w-w_i^0|^2.
\end{equation}
Here $w_i^0,{w_i^0}'$ are the centers of circles $C_i, C_i'$ correspondingly.
This can be satisfied for a special position of circles:
\begin{equation}
w_i^0 = {\lambda_i b_i - a_i\over \lambda_i - 1} \,, \qquad
{w_i^0}' = {\lambda_i a_i - b_i\over \lambda_i - 1} \,.
\label{*2}
\end{equation}
For such position of the circles their radii $\rho_i,\rho_i'$ become
equal. Thus, we have to compare (\ref{*1}) to the terms in
(\ref{action-takht}) for this special choice of the position
of circles. This is a simple exercise, for when the circles
$C_i$ are centered as prescribed by (\ref{*2}), the function
$\ln{|L_i'|}$ is constant along $C_i$ (proportional to $\ln{\rho_i}$),
and the calculation of all the contour integrals in 
(\ref{action-takht}) is straightforward. Performing this simple
calculation one finds that (\ref{*1}) and (\ref{action-takht})
are equal. 

A note is in order about the term involving $e^{2\phi}$ in the
``bulk'' part of (\ref{action-takht}). On-shell, this term is
a constant, proportional to $2g-2$. Our regularization procedure,
with the choice of the family of surfaces (\ref{surf}),
does not reproduce this constant. One can get this term correctly
by modifying the family of surfaces slightly. As is 
argued in \cite{D1D5}, the correct relation between the
Liouville field and the $z$ coordinate should involve the 
coupling to the boundary curvature $R$. Introducing such
a coupling one can get the term involving $e^{2\phi}$ correctly.
We shall not demonstrate this here. Instead, one can take another
point of view that any regularization procedure is unavoidably 
ambiguous. The Takhtajan-Zograf action fixes this ambiguity of adding
a constant to the action in a natural way.

The arguments presented show that the regularized volume computed using the above
``canonical'' family of regularizing surfaces is indeed given by
the Takhtajan-Zograf action. Note that in order to use the simple
family (\ref{surf}) of regularizing surfaces we had to place the
circles $C_i, C_i'$ at special positions. One can presumably avoid
doing this by modifying the regularizing family of surfaces appropriately,
but we shall not try to perform this more general calculation.
Note also that a part of the expression (\ref{*3}) that we obtained
on an intermediate stage of our calculation is exactly the
earlier result (\ref{action-reg}). Thus, one could also start
from the result (\ref{action-reg}) and add to it the ``bulk''
Liouville action plus a collection of contour integrals adjusted
so that the whole quantity renders the Liouville equations of motion
under the variation. The total action can again be shown to be
equal to the Takhtajan-Zograf action (\ref{action-takht}).

Having established that the regularized action for our spaces is
given by the Liouville action evaluated on the ``uniformizing'' 
Liouville field, we can discuss properties of the boundary CFT. 

\subsection{Stress-energy tensor.}
\label{ss:energy}

Recall, see, e.g., \cite{FS}, that the stress-energy tensor of a 
CFT can be thought of as a connection in a 
certain line bundle over the moduli
space, this connection being compatible with a metric on the
bundle, the role of the metric played by the partition function $Z$:
\begin{equation}
T = - Z^{-1} \partial Z,
\end{equation}
where the derivative is taken with respect to the moduli. The regularized
action $I$ is not yet (minus logarithm of) the partition 
function, the later can be obtained
from $I$, in the semi-classical approximation, by summing over all
solutions with the same conformal structure at infinity. However, it must
be possible to interpret these solutions as the semi-classical states 
of the CFT; thus, the derivative of the free energy $I$ with respect to
the moduli should have the
interpretation of the stress-energy tensor for these states. The
stress-energy tensor for each particular solution is also of direct
importance in the thermodynamical considerations of the next
subsection. It might also be of use in the problem of 
classification of solutions of (2+1)-gravity.

Thus, we would like to find the stress-energy tensor for each
solution. By putting the action $I$ in the form (\ref{action-takht})
we achieved this goal, for the dependence of (\ref{action-takht})
on the moduli was studied in \cite{Takht}. The
Theorem 1 of \cite{Takht}, applied to the action (\ref{action-takht}),
states that:
\begin{equation}
{1\over 2} {\partial I\over\partial\w_i} = - {l\over 16\pi G} c_i,
\label{theorem1}
\end{equation}
where $\w_i$ are the coordinates (\ref{coord-schottky}) on the
Schottky space $\Sch_g$, and $c_i$ are the coefficients in the
decomposition of the Liouville stress-energy tensor $T_w$, or,
the same, the Schwarzian derivative of the map $J^{-1}$, into
the basis of quadratic differentials $P_i$ introduced in the Appendix. 

We find the fact that the stress-energy tensor is given by
the Schwarzian derivative of the uniformization map extremely appealing, for
it has a nice CFT interpretation. Namely, in CFT one is used to the
fact that, because of the presence of the conformal anomaly, the stress-energy
tensor does not remain invariant under conformal transformations.
More precisely, it is only invariant under the ``global''
conformal transformations, that is the ones mapping the
complex sphere to itself, $\conf$ transformations. However, in
2d any change of the complex coordinate on the worldsheet by
an analytic function of it is a (locally) conformal transformation.
Because of the conformal anomaly, the stress-energy tensor is 
not invariant under such more general transformations, and transforms 
by acquiring a multiple of the Schwarzian derivative of the map.
These more general conformal transformations are the ones changing the
global topology of the worldsheet. Thus, one can interpret the
stress-energy tensor change under such a transformation as 
the Casimir energy ``acquired'' in the change of topology. The
well-known example of this phenomenon is the Casimir energy of
CFT on a cylinder, where it is given by the Schwarzian derivative of
the map from the plane to the torus. More interesting
example is given by the conformal dimension of the vacuum state
in cyclic orbifold theories, which turns out to be the
Schwarzian derivative of a map covering the cylinder $n$ times,
see \cite{Halpern}.

The boundary CFT we are discussing can also be thought as 
obtained by the orbifold construction. Indeed, all
our Riemann surfaces are obtained from the complex plane 
by discrete identifications. Thus, one could think of 
some ``mother'' conformal field theory living on the plane,
from which only certain states survive the identifications.
Those states then are used to construct the partition function on
a Riemann surface $X$. Then the vacuum state energy of the orbifolded
CFT must be given by the Schwarzian derivative from the plane
to $X$.  Equation (\ref{theorem1}) shows this to be indeed the case.

It would definitely be of interest to calculate the stress-energy
tensor, and parameters $c_i$ explicitly in some simple cases, 
as that of $g=2$. It would be even more interesting to find
a relation between parameters $c_i$ and the mass and angular momentum
parameters of the corresponding spacetimes. This would be interesting, 
for it would give an approach to the classification of solutions
of (2+1) gravity with negative cosmological constant. Indeed,
we saw that the corresponding Euclidean spaces are easy to classify
by their boundaries, and then the dependence of their free
energy $I$ on the moduli would allow one to find their
mass and angular momentum parameters. We hope to develop such a 
classification scheme in the future.

\subsection{Thermodynamics.}
\label{ss:therm}

We are now ready to study the thermodynamics of our spaces. The
discussion in this subsection is mostly qualitative, for we were
not able to explicitly calculate the action $I$ except in the
genus one case. However, one can see the general picture even 
without knowing the action explicitly.

As we mentioned before, the regularized action $I$ has the
meaning of the free energy of the corresponding space. As is
clear from the construction leading to (\ref{action-takht}),
it depends on the parameters of the Schottky groups generators,
that is, it is a function on the Schottky space. Thus, the
Schottky space $\Sch_g$ should be interpreted as the space of allowed
configurations (states) at genus $g$. As we saw
in the previous subsection, deriving $I$ with respect to the
coordinates $\w_i$ on the Schottky space, we get components
of the Liouville stress-energy tensor. This fact allows us to
write the first law of thermodynamics for our spaces.
Indeed, one has:
\begin{equation}
\delta I = \sum_{i=1}^{3g-3} {\partial I\over\partial\w_i} \delta\w_i =
\sum_{i=1}^{3g-3} \left ( - {l c_i\over 16\pi G} \right ) \delta\w_i,
\label{first-law}
\end{equation}
where we have used (\ref{theorem1}). Thus, the coordinates $\w_i$
on the Schottky space play the role of intensive parameters, that is, 
they are analogs of temperature, pressure. On the other hand, the 
parameters $c_i$ play the role of extensive variables, such as
energy, volume for usual systems. The equation (\ref{first-law})
is then just the first law of thermodynamics written in terms of the
free energy.

As we discussed in \ref{ss:lor}, our spaces have the
interpretation of ``Euclidean continuations'' of black hole
spacetimes. Such spacetimes have asymptotic regions, and thus 
are described by mass and angular momentum parameters for
each asymptotic region. In addition, there are parameters
describing the shape of the wormholes, hidden behind the horizons,
or the relative position of the horizons with respect to 
one another, see \cite{Brill} for more details. These
parameters are of direct physical significance, but, in
general, do not coincide with the parameters $c_i$, as well
as the thermodynamically conjugate quantities $w_i$ do not
coincide with temperatures, pressures etc. Thus, although the equation
(\ref{first-law}) has the interpretation of the first law,
it is not the first law in the form involving the variations
of the physical parameters. It would be quite desirable to
find a relation between $\w_i$ (or other known coordinates
on the Schottky and Teichmuller spaces) and the physical
parameters. One could then get a more informative form of the
first law. 

The problem of finding the physical parameters, such as
the mass and angular momentum, in terms of the parameters
$c_i$ is also directly related to the problem of finding
the entropy of our spaces. Indeed, recall that the entropy
of a ``usual'' thermodynamical system can be obtained
from the free energy as: $S = \beta E - I$;
there is no volume or pressure in this formula. Recall that
all parameters describing our spaces can be divided into
parameters describing the asymptotic regions, and the 
parameters specifying the ``internal'' geometry. The later
ones are the analogs of pressure in the ``usual'' system. Thus, it is
natural to define the entropy of a space under consideration
as, schematically
\begin{equation}
S = \sum_{\vbox{\hbox{\rm asymptotic}\hbox{\,\,\,\,\,regions}}} 
\left( \beta E + \Phi J \right) - I,
\end{equation}
where $\beta,\Phi$ are the conjugate variables to the energy $E$ and
the angular momentum $J$; the sum is taken over all asymptotic
regions. It would be quite desirable to try to define the physical
parameters $E,J$ and their conjugate ``intrinsically'', using only the geometrical
properties of the Schottky (Teichmuller) space. This would
then allow one to define the entropy function on the Teichmuller
space. One expects this function to be equal to the quarter of
the sum of the circumferences of all the horizons. Work is in
progress to check whether this is indeed the case.

The thermodynamical stability of our spaces can also be discussed
from the same general viewpoint. Recall that, in order to find
whether a particular state is thermodynamically stable, one 
finds the matrix of specific heats at that state. If all
of its eigenvalues are positive, the state is stable. To find
the specific heat matrix for our spaces, we should derive the
extensive parameters $c_i$ (``energies'') with respect to 
the intensive parameters $\w_i$ (``temperatures''). 
This problem was solved in \cite{Takht}. The Theorem 2 of this
reference states:
\begin{equation}
{\partial c_i\over\partial\bar{\w}_j} = -{1\over2} \langle
{\partial\over\partial\w_i}\, , \, {\partial\over\partial\w_j} \rangle,
\label{theorem2}
\end{equation}
where $\langle\cdot\, , \,\cdot\rangle$ is the Weyl-Peterson
metric on the Schottky (Teichmuller) space, reviewed in the
Appendix. Thus, to analyze the thermodynamical stability of
a particular state, one should just analyze the eigenvalues of
the Weyl-Peterson metric at the corresponding point of the 
Schottky space. In general one should not expect the spaces
considered to be ``absolutely'' stable, in the sense that
all specific heats are positive. Indeed, the black hole spacetimes
are, except in the simplest case of BTZ black hole, collapsing. 
One should, however, expect the 
asymptotic regions to be stable. Thus, at least the
specific heats corresponding to the asymptotic regions 
must be positive.

Let us now turn to the question of phase transitions. As is
well known, see \cite{HP}, in asymptotically AdS spaces one
has a phase transition (AdS at low temperatures) vs. (BH at high
temperatures). In the context of AdS/CFT correspondence, this
phase transition was interpreted in \cite{Witten-AdS}
as the phase transition in the boundary CFT. In the case
of 2+1 dimensions, both periodically identified in the time
direction AdS and the Euclidean black hole have the topology
of the solid torus. Thus, the boundary has genus $g=1$.
It is interesting whether one also gets phase transitions
in the higher genus case. A simple argument shows that this
must indeed be the case. Let us recall how one finds that
there is a phase transition in the genus one case. Calculating
the free energies of AdS and BTZ spaces, one finds that the
free energy of AdS is the smallest of the two at small 
temperatures, while the free energy of BTZ is the smallest at
large temperatures. The point where the two free energies are
equal is the phase transition point. It is the point when the
horizon radius of BTZ black hole is equal to the curvature radius
of AdS: $r_+ = l$. One can observe that AdS and the black hole
spaces of the same temperature define the same conformal structure
on the boundary. However, they realize different points in the
Schottky space. As we discussed in \ref{ss:euclid}, the
Schottky space of the torus can be realized as the strip of
the upper half-plane. Then AdS space and BTZ black hole of the
same temperature are represented by two points on the
imaginary axes related by the modular transformation
$\tau\to-1/\tau$. The point where the phase transition
occurs is the point where the free energies are equal, in other words,
it is the point which is invariant under the above modular
transformation. Thus, it is at points of the Schottky space 
which are invariant under some modular transformations that
the free energies of different solutions are equal. At this
point all such solutions are equally important; this is the
point of a phase transition. Thus, one should
expect phase transitions to occur at points of the Schottky space
which are invariant under some modular transformations. These
are exactly the points where the corresponding Riemann space
has conical singularities. Such points exist not only in the genus
one case, but also in the higher genus cases, where their pattern
becomes much more complicated. It would be interesting to find
these points for the case of $g=2$, and find which
phase transitions they correspond to. 

\subsection{Boundary CFT}
\label{ss:cft}

We are now ready to study the boundary CFT partition function. The rules
of bulk/boundary correspondence tell us that the semi-classical approximation
to the partition function is given by the sum over all classical 
solutions which define the same conformal structure at the
boundary. We saw that different classical solutions correspond
to different points in the Schottky space. Two points that are
related by a modular transformation correspond to the
same conformal structure. Thus, to construct 
the partition function we have to take a sum over the action of the modular
group on $\Sch_g$ of the exponentials of (minus) the regularized action:
\begin{equation}
Z(\w) =\sum_{\sigma\in{\rm Mod}(\S_g)} e^{- I[\sigma(\w)]}.
\label{part}
\end{equation}
Here the sum is taken over the elements $\sigma$ of the set ${\rm Mod}(\Sch_g)$
of modular transformations of $\Sch_g$. Recall that ${\rm Mod}(\Sch_g)$
is the quotient of the group ${\rm Mod}(\Gamma)$ of all modular
transformations by the normal subgroup consisting of those elements
that send the group $N\subset\Gamma$ into itself. In (\ref{part}),
$\w$ stands for the set of coordinates on the Schottky space
$\Sch_g$. Thus, assuming the sum in (\ref{part}) converges, the partition 
function is a function on the Schottky space, which is by 
construction invariant under the modular transformations. It, therefore,
descends to a function on the Riemann space. We note that
the semi-classical approximation (\ref{part}) to the
partition function was studied, in the genus one case, in
papers \cite{MS,Mano,Farey}.

It is a good place to emphasize the role played in the whole
construction by the Schottky space. As we see,
the Schottky space is more important in our scheme
than the Teichmuller space. It is different points in the Schottky
space that correspond to different classical solutions. Points
in the Teichmuller space that cover a particular point in $\Sch_g$
just correspond to different choices of the basis of the generators
$\a$ of $\pi_1(X)$. As we
saw, the regularized action is a function on the Schottky space.
It can be, of course, thought of as a function on Teichmuller
space, but it will be invariant under the elements of ${\rm Mod}(\Gamma)$
that send $N\subset\Gamma$ to itself. A good analogy here is that a with
gauge symmetry: the ``gauge invariant'' quantities only depend on a point
in $\Sch_g$, not on a point in its covering space $T_g$. This analogy also
gives another point of view on why we sum in (\ref{part}) over
the modular transformations on $\Sch_g$, not on $T_g$: it is usual
in the path integral for a system with a gauge symmetry to sum 
only over the equivalence classes. To summarize, the
Schottky space is more fundamental in our construction. In a sense,
the Schottky space is the ``reduced phase space'', while the
Teichmuller space is the phase space where the gauge symmetry
acts. 

What is the CFT to whose partition function the expression 
(\ref{part}) is an approximation? One possibility, suggested rather
strongly by the role played in our story by the Liouville theory,
is that this CFT is nothing else but the quantum Liouville theory. 
This theory is still far from being
understood completely, despite some recent progress \cite{Q-Liouv}
on the subject, but the recent results seem to suggest that the theory
does make sense quantum mechanically. If so, it would be extremely
interesting to see in what regime the exact Liouville partition
function is approximated by (\ref{part}). Another possibility
is that Liouville theory does not make sense quantum mechanically,
or is not enough to account for all of the degrees of freedom
expected to be found in (2+1)-dimensional quantum gravity. This would mean
that there is no quantum theory of pure gravity
in 2+1 dimensions (with negative cosmological constant). This 
possibility is advocated by some string theorists, see, e.g., 
\cite{Martinec}. If this turns out to be the case, one would
only be able to interpret (\ref{part}) as an approximation to
the partition function of the CFT's that arise in the context of
AdS/CFT correspondence from string/M theory compactified on
backgrounds containing AdS${}_3$. These CFT's are rather
complicated, containing many fields. The expression (\ref{part})
would then be interpreted as (an approximation to) the full
generating functional, in which all the sources except for the
stress-energy tensor source are put to zero.

We conclude this subsection by mentioning the
possible orbifold interpretation of the CFT on higher genus
surfaces. As we already discussed in \ref{ss:energy}, the semi-classical
states of our theory (classical solutions) might be thought of
as obtained from some ``mother'' conformal field theory on the
plane by selecting states invariant under the discrete group
$\Sigma$. The experience with orbifolds tells that one should
expect the stress-energy for such states to be given by
the Schwarzian derivative of the covering map. As we saw,
the states considered have this property. It would be very
interesting to develop this orbifold interpretation in more
details.

\section{Outlook}
\label{sec:out}

In this section we briefly discuss what our results imply for the
problem of quantization of 2+1 gravity.
Gravity in 2+1 dimensions is an extremely rich and beautiful theory. 
In the simplest case, when no matter is present and the space is compact,
the theory is topological - its only degrees of freedom are global, topological 
properties of the underlying manifold. In this case it can be successfully
quantized either canonically \cite{Witten:2+1} or 
via recurse to Chern-Simons theory and its path integral
\cite{Witten:CS}. This relates 2+1 gravity to such vast subjects as topological 
quantum field theory, knot theory, quantum groups. More complicated
situations arise when one allows matter sources to be present. In fact,
in this case one can get quite a non-trivial dynamics, in spite of the
fact that there are no local degrees of freedom of the gravitational
field itself. An example
of such non-trivial dynamics, for the case when matter is just a collection of point 
particles, was exhibited in \cite{Hooft:particles}.

Another rich case is that
of a negative cosmological constant, both with and without
matter. In this case the space is
not compact; one has asymptotic infinity with its rich 
structure. Simple arguments \cite{Brown-H} 
based on semi-classical considerations show that quantum 
gravity in this case must be describable by some CFT living on the
boundary. The old results of \cite{Ass-H} suggest that this 
CFT is the to-be-constructed quantum Liouville theory. The results of \cite{Ass-H},
however, pertain only to the simplest boundary topology, that
of $S^1\times\R$ in the Lorentzian case, or $S^2$ in the Euclidean regime.
Our results for the general genus demonstrate it more convincingly
that the boundary CFT must indeed be the Liouville theory.

We have already mentioned the point of view on the Schottky
space as the ``reduced phase space'', the ``full phase space''
being the Teichmuller space. In fact, this is more than an
analogy. Asymptotic degrees of freedom of gravity with negative 
cosmological constant are the conformal structures one can
put at infinity; thus, the Teichmuller space can indeed be 
thought of as the space of states of asymptotically AdS gravity.
Moreover, both Teichmuller and Schottky spaces are symplectic
manifolds, and thus are indeed phase spaces; the symplectic two-form is
given by the imaginary part of the Weyl-Peterson hermitian metric. 
Thus, one can study the problem of quantization of these phase spaces. 
There is a growing body of evidence, see \cite{Verlinde} and more
recent papers \cite{QT1,QT2}, that 
this quantum theory is nothing but the Liouville theory. What
we found by studying the bulk/boundary correspondence also suggests
that the boundary theory is the Liouville theory. This convergence
of two different lines of thought is impressing.

If the boundary CFT is indeed the Liouville theory, then 
the bulk/boundary correspondence provides us with an interesting 
perspective on it. As is clear from our construction, the semi-classical 
states of the boundary CFT have the interpretation of classical solutions.
The question then arises what spaces (spacetimes) in the bulk, if any, 
correspond to the ``elementary'' quantum states on the boundary. In fact,
one may reverse the question and try to construct the quantum
Liouville states as the ones corresponding to a particular nice set of 
spaces in the bulk. Such an approach to the quantum Liouville would be
very much in the spirit of holography; it seems to be worth pursuing.

\section{Acknowledgments}

I would like to thank A.\ Ashtekar, J.\ Baez, D.\ Brill, J.\ David, A.\ Hashimoto,
K.\ Skenderis, J.\ Maldacena for discussions. I am grateful to L.\ Takhtajan and 
P.\ Zograf for correspondence. I am especially grateful to 
G.\ Horowitz for getting me interested in the subject as well as for
discussions. This work was supported in part by NSF grant PHY95-07065.

\appendix

\section{Teichmuller theory}
\label{sec:teich}

Here we review some well-known facts on Teichmuller theory. Our presentation
closely follows that of \cite{Takht}.

The space of possible conformal structures on a genus $g$ Riemann
surface (the Riemann space) has a complicated structure, and it
was the idea of Teichmuller to consider another, simpler space, out
of which the Riemann space can be obtained as the quotient by a group
of discrete transformations. The Teichmuller space is usually defined
as the space of equivalence classes of quasiconformal mappings from a
fixed base Riemann surface, see, e.g., \cite{Univalent}. However,
in practice, it is more convenient to uniformize Riemann surfaces
by Fuchsian groups, and consider the space of equivalence classes
of Fuchsian groups obtained from a fixed, base Fuchsian group
by quasiconformal transformations. The Teichmuller space so defined
comes equipped with a coordinate system: these are the so-called
Fricke coordinates, in which the Teichmuller space is a region
in $\R^{6g-6}$. 

Thus, let $\Gamma$ be a Fuchsian group uniformizing a marked
Riemann surface $X$ of genus $g>1$. This Fuchsian 
group will serve as the base point. The
Beltrami differential for $\Gamma$ is a complex-valued, bounded
function $\mu$ on the unit disk $H$, having the property: 
$\mu(\gamma z) \overline{\gamma'(z)}/\gamma'(z) = \mu(z)$
for all $z\in H, \gamma\in\Gamma$. The vector space of all
such functions will be denoted by $B(\Gamma)$. For every $\mu\in D(\Gamma)$,
where $D(\Gamma)$ is the open unit ball (in the sense of the sup-norm)
in $B(\Gamma)$, the Beltrami differential equation:
\begin{equation}
f^\mu_{\bar{z}}(z) = \mu(z) f^\mu_z(z)
\end{equation}
has a unique (up to transformations from ${\rm PSL}(2,\R)$) solution
in the class of quasiconformal maps of $H$ on itself. For every solution
$f^\mu$, the group $\Gamma^\mu=f^\mu \Gamma (f^\mu)^{-1}$ is a Fuchsian
group uniformizing the Riemann surface that is obtained from $X$ as the
result of the quasiconformal map $f^\mu$. Thus, every element $\mu\in D(\Gamma)$
gives a representation of $\Gamma$ in the group ${\rm PSL}(2,\R)$:
$\gamma \to f^\mu \gamma (f^\mu)^{-1}$. The set of equivalence
classes of such representations, where the equivalence is by
inner automorphisms from of ${\rm PSL}(2,\R)$, 
is called the Teichmuller space for $\Gamma$ and is denoted by 
$T(\Gamma)$. Let us note that choosing a different base point 
one obtains the same Teichmuller space. Thus, when the choice of 
the base point is not important, we will denote the Teichmuller
space by $T_g$.

Thus, every Beltrami differential $\mu\in D(\Gamma)$ corresponds to
a point in the Teichmuller space $T(\Gamma)$. Let us denote the
corresponding map by $\Phi$:
$$\Phi: D(\Gamma) \to T(\Gamma).$$
Note that every Fuchsian group can be characterized by $6g$ parameters
minus 3 relations (\ref{Fuchs-rel}). The equivalence classes are
thus characterized by $6g-6$ parameters, this gives the so-called
Fricke coordinates on the Teichmuller space. The Teichmuller modular 
group ${\rm Mod}(\Gamma)$ acts on $T(\Gamma)$ by discrete transformations, 
and the corresponding quotient space is the Riemann space:
$R(\Gamma)=T(\Gamma)/{\rm Mod}(\Gamma)$.

Let us now introduce the so-called Weyl-Peterson metric on $T(\Gamma)$.
First, we need a pairing between Beltrami differentials and 
holomorphic quadratic differentials. Let us define
the space $\Q(\Gamma)$ of holomorphic quadratic differentials
for group $\Gamma$ to be the space of holomorphic 
functions $q(z)$ on $H$ satisfying $q(\gamma z)\gamma'^2(z)=q(z)$
for all $\gamma\in\Gamma, z\in H$. The Riemann-Roch theorem
implies that ${\rm dim}_\C \Q(\Gamma)= 3g-3$. We have a pairing:
\begin{equation}
(\mu,q) = \int\!\!\!\!\int_{H/\Gamma} \mu(z) q(z) 
{|dz\wedge d\bar{z}|\over 2},
\label{pairing}
\end{equation}
which is invariantly defined because of the transformation properties
of $\mu, q$. Let $N(\Gamma)\subset B(\Gamma)$ be the kernel of the
above pairing. Then in every class $B(\Gamma)/N(\Gamma)$ there exists
a representative of the form $({\rm Im}z)^2 \overline{q(z)}, q\in \Q(\Gamma)$.
Such Beltrami differentials are called harmonic. The space of harmonic
Beltrami differentials will be denoted by ${\cal H}(\Gamma)$.

The kernel of the differential $d\Phi$ of the map $\Phi:D(\Gamma)\to T(\Gamma)$
at point $\Phi(0)$ coincides with $N(\Gamma)$. Thus, the space ${\cal H}(\Gamma)$
can be identified with the tangent space to $T(\Gamma)$ at $\Phi(0)$. The
pairing (\ref{pairing}) allows then to think of $\Q(\Gamma)$ as the
cotangent space to $T(\Gamma)$ at $\Phi(0)$. The tangent and cotangent
spaces at other points of $T(\Gamma)$ can be realized as ${\cal H}(\Gamma^\mu),
\Q(\Gamma^\mu)$. 

One can now define a hermitian metric on $T(\Gamma)$:
\begin{equation}
\langle \mu_1,\mu_2 \rangle = (\mu_1,({\rm Im}z)^{-2} \bar{\mu}_2),
\qquad \mu_1,\mu_2\in {\cal H}(\Gamma^\mu).
\label{weyl-pet}
\end{equation}
This is the Weyl-Peterson metric. It is Kahler and invariant
under ${\rm Mod}(\Gamma)$.

Let us now discuss a relation between the Teichmuller and
Schottky spaces. As we discussed in \ref{ss:euclid},
given a marked Riemann surface $X$, or, equivalently, a
marked Fuchsian group $\Gamma$, one can construct the
corresponding normalized marked Schottky group $\Sigma$.
Thus, we have a map:
$$\Psi: T_g\to \Sch_g\subset \C^{3g-3}.$$
The Teichmuller space is the universal covering
space for $\Sch_g$: the group of automorphisms of $\Psi$
is isomorphic to the fundamental group $\pi_1(\Sch_g)$,
and consists of those elements of ${\rm Mod}(\Gamma)$ that
send the subgroup $N\subset\Gamma$ into itself.

Because $\Sch_g$ is realized as a subspace of $\C^{3g-3}$,
one has on it a preferred set of vector fields 
$\partial/\partial\w_i$, where $\w_i, i=1,\ldots,3g-3$ are
the coordinates (\ref{coord-schottky}) on the Schottky space.
Using the map $\Psi$, they can be lifted to vector fields
on $T_g$. The tangent bundle over $T_g$ is the space of 
holomorphic Beltrami differentials; one thus gets the
basis of vector fields $\nu_i^\mu$ on $T_g$ that correspond
to $\partial/\partial\w_i$ on $\Sch_g$:
\begin{equation}
d\Psi_{\Phi(\mu)} (\nu_i^\mu) = {\partial\over\partial\w_i},
\qquad \nu_i^\mu\in{\cal H}(\Gamma^\mu), i=1,\ldots,3g-3.
\end{equation} 
One can then introduce the dual basis of quadratic differentials
$p_i^\mu, p_i^\mu\in\Q(\Gamma^\mu)$. Let us also introduce the
corresponding quadratic differentials $P_i^\mu$ on $\S$:
\begin{equation}
P_i^\mu = p_i^\mu \circ J_\mu^{-1} (J_\mu^{-1})'^2,
\label{basis-P}
\end{equation}
where $J_\mu$ is the map from the hyperbolic plane $H$ to the
region of discontinuity $\Omega_\mu$ of the Schottky group
$\Sigma^\mu$ corresponding to $\Gamma^\mu$. The quadratic
differentials $P_i^\mu$ are holomorphic functions of $\Omega_\mu$,
automorphic with respect to $\Sigma^\mu$. The Schwarzian derivative
of $J_\mu^{-1}$, which is a holomorphic quadratic differential
on $\Omega_\mu$, can be decomposed into the basis of $P_i^\mu$:
\begin{equation}
\{ J_\mu^{-1}(w) ; w \} = \sum_{i=1}^{3g-3} c_i^\mu P_i^\mu(w).
\label{accessory}
\end{equation}
The quantities $c_i$ are functions on the Schottky space: 
$c_i^\mu$ depends on the
parameters of the Schottky group $\Sigma^\mu$. As we show in 
\ref{ss:therm}, they play the role of the extensive
thermodynamical parameters for our spaces, while the coordinates
$\w_i$ on the Schottky space are the intensive parameters.

\end{document}